\documentclass[conference]{IEEEtran}
\usepackage{cite}
\usepackage{amsmath,amssymb,amsfonts}
\usepackage{algorithmic}
\usepackage{graphicx}
\usepackage{textcomp}
\usepackage{xcolor}
\usepackage{float}
\usepackage{hyperref}
\usepackage{quantikz}
\usepackage{fancyvrb}
\hypersetup{colorlinks=true, % colorise les liens
breaklinks=true, % retours à la ligne pour les liens trop longs
urlcolor= black, % couleur des hyperliens
linkcolor= black,% couleur des liens internes aux documents 
citecolor= black % couleur des liens vers la biblio
} % comment or uncomment to colorbox the hyperref links !

\usepackage{cleveref} 
\Crefname{figure}{Fig.}{}
\Crefname{equation}{}{}
\Crefname{paragraph}{paragraph}{}

\usepackage{braket}
\newcommand\ketbra[1]{\ensuremath{ \ket{#1} \bra{#1} }}
\usepackage{scalerel,stackengine}
\stackMath
\newcommand\reallywidehat[1]{%
\savestack{\tmpbox}{\stretchto{%
  \scaleto{%
    \scalerel*[\widthof{\ensuremath{#1}}]{\kern-.6pt\bigwedge\kern-.6pt}%
    {\rule[-\textheight/2]{1ex}{\textheight}}%WIDTH-LIMITED BIG WEDGE
  }{\textheight}% 
}{0.5ex}}%
\stackon[1pt]{#1}{\tmpbox}%
}

\usepackage{acro}
\DeclareAcronym{neeqma}{short=\textsc{NEEQMA}, long=Numerical Error Extraction by Quantum Measurement Algorithm}
\DeclareAcronym{ftqc}{short=\textsc{FTQC}, long=Fault Tolerant Quantum Computing}
\DeclareAcronym{isq}{short=\textsc{ISQ}, long=Intermediate Scale Quantum Computing}
\DeclareAcronym{qpe}{short=\textsc{QPE}, long=Quantum Phase Estimation}
\DeclareAcronym{hhl}{short=\textsc{HHL}, long=Harrow–Hassidim–Lloyd algorithm}
\DeclareAcronym{qsp}{short=\textsc{QSP}, long=Quantum Signal Processing}
\DeclareAcronym{qsvt}{short=\textsc{QSVT}, long=Quantum Singular Value Transform}
\DeclareAcronym{lcu}{short=\textsc{LCU}, long=Linear Combination of Unitaries}
\DeclareAcronym{pf}{short=\textsc{PF}, long=Product Formula}
\DeclareAcronym{qpu}{short=\textsc{QPU}, long=Quantum Processing Unit}
\DeclareAcronym{hf}{short=\textsc{HF}, long=Hartree-Fock}
\DeclareAcronym{bch}{short=\textsc{BCH}, long=Baker-Campbell-Hausdorff}
\DeclareAcronym{mpf}{short = MPF, long = Multi Product Formula}
\DeclareAcronym{dmpf}{short = DMPF, long = Dynamical Multi Product Formula}
\DeclareAcronym{ld}{short = LD, long = Limited Development}
\DeclareAcronym{ld1}{short = LD1, long = Limited Development n:1}
\DeclareAcronym{ld2}{short = LD2, long = Limited Development n:2}
\DeclareAcronym{zne}{short = ZNE, long = Zero-Noise Extrapolation}
\DeclareAcronym{ts}{short = TS, long = Taylor Series}

\usepackage[colorinlistoftodos, textsize=small]{todonotes}

\usepackage{tikz}
\usetikzlibrary{positioning,arrows,calc,math,angles,quotes}
\usepackage{multirow}

\def\BibTeX{{\rm B\kern-.05em{\sc i\kern-.025em b}\kern-.08em
    T\kern-.1667em\lower.7ex\hbox{E}\kern-.125emX}}
\begin{document}

%%%%%%%%%%%%%%%%%%%%%%%%%%%%%%%%%%%%%%%%%%%%%%%%%%%%%%%%
%%% Core

\title{\acl{neeqma}}

\def\orgadep{CEA, List, F-91120}%Laboratoire Intégration des Systèmes et Technologies (LIST)} % dept. name of organization (of Aff.)
\def\orga{CEA}%Commissariat à l'Energie Atomique et aux Energies Alternatives (CEA)} % name of organization (of Aff.)
\def\loc{Palaiseau, France} % City, Country
\def\univ{Université Paris-Saclay}

\IEEEoverridecommandlockouts

\author{\IEEEauthorblockN{Clément RONFAUT}%$^*$}
\IEEEauthorblockA{\textit{\univ} \\
\textit{\orgadep}\\
\loc \\
0009-0006-4714-0731}
\and
\IEEEauthorblockN{Robin OLLIVE}%$^*$}
\IEEEauthorblockA{\textit{\univ} \\
\textit{\orgadep}\\%\orga}\\
\loc \\
0009-0006-7539-363X} % email address or ORCID
\and
\IEEEauthorblockN{Stéphane LOUISE}
\IEEEauthorblockA{\textit{\univ} \\
\textit{\orgadep}\\
\loc \\
0000-0003-4604-6453}
% \thanks{$^*$These authors contributed equally to this work.}
}

\maketitle

\begin{abstract}
  In quantum computing, some of the main quantum operations (\textit{i.e.} gates) cannot be simply implemented and require an approximation.
  Major quantum algorithm routines allow the implementation of these specific quantum gates by combining basic quantum circuits with an iterative structure.
  % Convergence parameters are associated with the number of repetitions of the basic circuit pattern.
  In this structure, the number of repetitions of the basic circuit pattern is associated to convergence parameters.
  % This iterative structure behaves similarly to function approximation by series expansion: the higher the 'truncation' order, the better the target gate (\textit{i.e.} operation) approximation.
  The asymptotic convergence of the gate error with respect to the number of basic pattern repetitions is known and is referred to as the query complexity.
  % Similarly, the gate complexity indicates the asymptotic convergence with respect to the number of quantum gates in the circuit.
  The underlying convergence law is bounded, but often lacks explicit constants for specific problem instances. % it is not always explicitly stated.
  Upper bounds are generally too pessimistic to be useful in practice.
  The actual convergence law contains constants that depend on the joint properties of the matrix encoded by the query and the initial state vector, which are difficult to compute classically.

  This paper proposes a strategy to study this convergence law directly from \ac{qpu} output.
  Given a convergence law, this protocol extracts the numerical value of the associated constants from the gate (\textit{i.e.} operation) approximation at different accuracies (\textit{i.e.} convergence parameter) constructed directly on the \ac{qpu}.
  This protocol is called \ac{neeqma}.
  \ac{neeqma} concepts are tested on specific instances of \ac{qsp} and Hamiltonian Simulation by Trotterization.
  Knowing the convergence constants associated with a semi-norm of the error allows for selecting the smallest convergence parameters that enable reaching the required gate approximation accuracy, thereby satisfying the quantum algorithm's requirements.
\end{abstract}

\begin{IEEEkeywords}
Trotterization, Approximation, Convergence, %complexity, 
\acf{qsp}, \acf{neeqma}
\end{IEEEkeywords}

\acresetall
\section{Introduction}
Some of the main quantum algorithms cannot be simply implemented and require an approximation that converges towards the solution depending on convergence parameters.
This iterative structure behaves similarly to function approximation by series expansion: the higher the 'truncation' order, the better the target gate (\textit{i.e.} operation) approximation.
Since reducing the approximation error comes with the cost of increasing the number of gates, it is of great matter to know explicitly the error law in order to choose these parameter values accordingly.
It is very difficult classically since the convergence law contains constants that depend on the joint properties of the matrix encoded by the query and the initial state vector.
Our contribution is to provide a quantum routine \ac{neeqma} capable of outputting a semi-norm of the gate error for a convergence parameter value.

One of the first algorithms proposed for quantum computers is to model the time evolution of quantum systems \cite{feynman_simulating_1982}.
Mathematically, it consists in applying the Hamiltonian simulation, a unitary matrix, to the system's initial state vector \cite{lloyd_universal_1996} \cite[section 4.7]{nielsen_quantum_2010}.
To this end, the proposed method consists of combining simpler systems' Hamiltonian simulation (known as summands) using the quantum routine called \ac{pf} \cite[Section 2.3]{childs_theory_2021} \cite{trotter_product_nodate, suzuki_generalized_1976, low_well-conditioned_2019, zhuk_trotter_2024}.
The simplest \ac{pf} consists of repeating the simpler system Hamiltonian simulation following the Lie-Trotter formula \cite[Section 3.1]{childs_theory_2021}, which is an approximation that converges toward the Hamiltonian simulation of the complete system.
Its error model can be derived using the \ac{bch} formula \cite[Section 1.1]{childs_theory_2021} \cite{lloyd_universal_1996}. %and a Taylor expansion \cite{childs_theory_2021}.
The convergence parameter that controls this convergence is called the Trotter number $n$ \cite{childs_theory_2021}.

% More recently, a routine named \ac{hhl} was proposed to block encode a function on the eigenvalues of a matrix.
% It uses the exact Hamiltonian simulation of the matrix as the initial query.
% A \ac{qpe} diagonalize the matrix, then quantum arithmetic is used to apply the function of interest on the eigenvalues.
% The images of the function are amplitude encoded on the block encoding control qubit, before the other gates are uncomputed to go back to the initial basis.
% Focusing on the error brought by the \ac{qpe} part, it is easy to understand that the higher the number of digits used to express the eigenvalues, the smaller the error on the inputs of the evaluated function and so on the final block encoding.
% The query complexity, which is the number of query calls required to achieve a given error, is equal to the number of ancilla qubits used to express the eigenvalues.
% As for the other routines, the coefficient of the complexity formula depends on the initial state vector on which the algorithm is applied. % Different state vector overlap with different eigenvalues, and the needed number of qubits to express the one with the larger number of decimal places can change.
%  Additionally, it depends on the quantum arithmetic implementation of the function in question.

Another major quantum routine derived from the Hamiltonian simulation is the \ac{qpe} algorithm \cite{kitaev_quantum_1995}.
It produces the spectral decomposition of the time evolved Hamiltonian, where the eigenstates are intricated with the binary expression of the corresponding eigenvalue\footnote{More accurately, it reads the eigenphase which are related to the eigenvalue by: $ \widehat{H_{P}} = \sum_{j} \lambda_{j} \ketbra{\lambda_{j}} $ therefore $ e^{i t \widehat{H_{p}}} = \sum_{j} e^{i t \lambda_{j}} \ketbra{\lambda_{j}} $.} in another quantum register.
The error of the spectral decomposition is directly linked to the number of qubits $m$ in the eigenvalue readout register.
It is also indirectly linked to the ability to construct the Hamiltonian simulation of a specific Hamiltonian $\widehat{H_{p}}$ at a specific time: $ t \geq \frac{2 \pi}{||\widehat{H_{p}}||} \times 2^{m} $ with an error on the ground-state eigenvalue smaller than $\varepsilon_{\textit{QPE}}$ and $||\widehat{H_{p}}||$ the spectral norm of $\widehat{H_{p}}$.
It has been demonstrated by \cite{sennane_robustness_2026} that the maximum Hamiltonian Simulation time only depends on the accuracy $\varepsilon_{\textit{QPE}}$: $ t \geq \frac{2 \pi}{\varepsilon_{\textit{QPE}}} $.
The number of readout qubits $m$ is a \ac{qpe} convergence parameter. %  ; comp(qc) = comp(QFT(m)) + m * comp(query)
If a \ac{pf} constructs the Hamiltonian simulation, the required number of repetitions to reach a small enough error is a second convergence parameter.
Interestingly, \ac{qpe} is a subroutine of the larger \ac{hhl} that enables block encoding of normalized functions on matrices' eigenvalues \cite{harrow_quantum_2009}.

A breakthrough in quantum computing was brought with the concept of qubitization \cite{low_hamiltonian_2019}, which allows the extension of the \ac{qsp} method to many-qubit Hamiltonians and its generalization to \ac{qsvt} for general matrices\cite{gilyen_quantum_2019, martyn_grand_2021}.
Similar to \ac{hhl}, it permits the construction of a block encoding of a matrix function \cite[Section II.E]{martyn_grand_2021}.
It is based on a different family of queries, known as signal operators, that are notably obtained from the matrix block encoding.
This technique block encodes a Chebyshev %or Laurent 
polynomial approximation of a function whose degree equals the number of query repetitions.
The polynomial coefficients are implemented through the phase angles, the parameters of the signal processing operator.
The signal processing operators are inserted between the signal operators.
The error in the matrix function block encoding varies in accordance with the error of the polynomial approximation of the target function evaluated on the eigenvalues.
It allows one to study the \ac{qsp} query complexity (for a given function) by knowing the convergence properties of a polynomial that approximates the function with a bounded error \cite{low_quantum_2017, martyn_grand_2021}.

\begin{table}[tb]
  \caption{Quantum Algorithm and Routine with associated Convergence Parameters and Error Models.}
  \center
  % \resizebox{\linewidth}{!}{
      \begin{tabular}{|c|c|c|}%c|}
		  \hline
		  \textbf{Routine} & \textbf{Convergence} & \textbf{Convergence} \\ % & \textbf{Ancilla} \\
      \textbf{/Algorithm} & \textbf{Parameter} & \textbf{low/Error Model} \\ % & \textbf{qubit(s)} \\
      \hline
      Lie-Trotter & Trotter number $n$ & $ \mathrm{Err}(n) $ : \Cref{eq_error_model_trotter} \\ % & $0$ \\
		  \hline
      \acs{qpe} & readout qubits $m$ & $ m = \lceil \log_{2}(\frac{||\widehat{H_{p}}||}{\varepsilon_{\textit{QPE}}}) \rceil $ \\ % & $n$ \\
      \hline
      \acs{qsp} & polynomial order $d$ & $ \mathrm{Err}(d) $ : \Cref{eq_error_model_qsp} \\ % & $1$ \\
      \hline
      \hline
      $ \mathrm{f}(x) $ & approx order $d$ & $ \varepsilon_{x}(d) = | \mathrm{f}(x) - \mathrm{poly}(d, x) | $ \\ % & / \\
      \hline
	\end{tabular} %}
  \label{table_convergence}
\end{table}

\begin{figure}[tb]
\begin{center}
\resizebox{\linewidth}{!}{\includegraphics{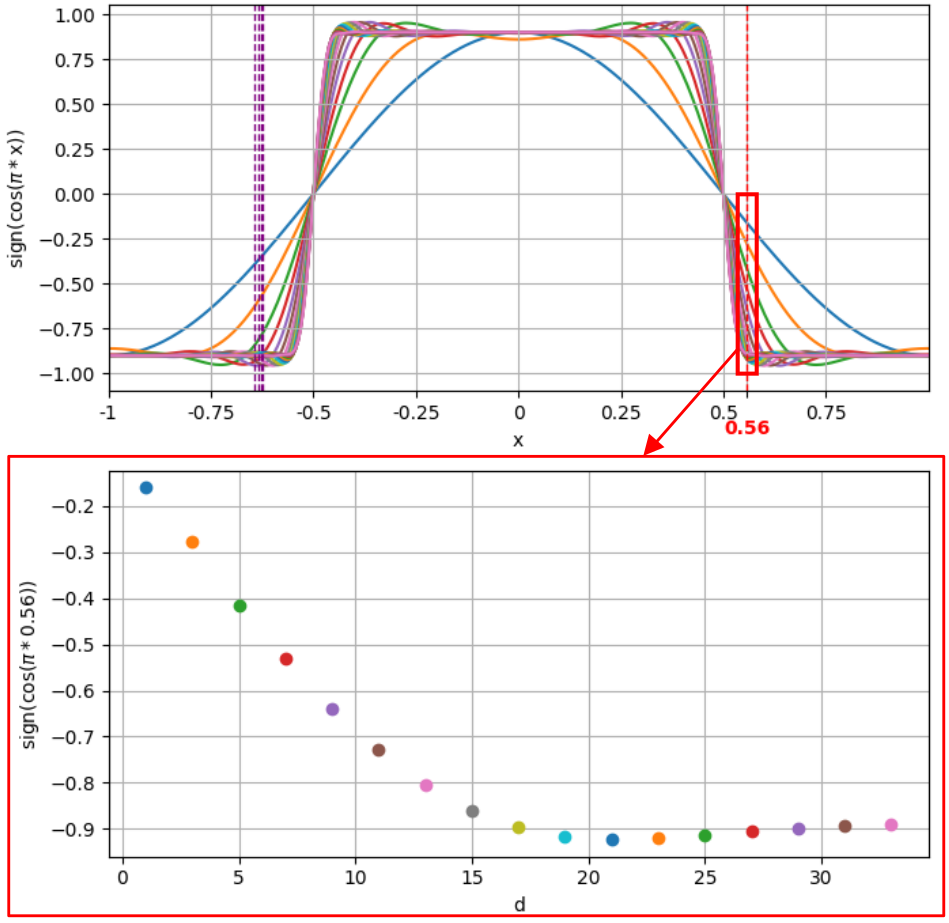}}
\end{center}
\caption{$ \mathrm{sign} \circ \cos(\pi x) $ function polynomial approximations at different orders $d$. The top graph represents the function approximations evaluated at many values in the $[ -1, 1 ]$ interval. It shows that the error fluctuates depending on the evaluation point. The bottom curve shows how the approximations converge as $d$ increases for a chosen value equals to $ x = 0.56 $ chosen for illustrative purposes; the associated error is a scalar.
The purple vertical lines correspond to the eigenvalues associated with the largest $\alpha$ values of \Cref{section_qsp} experiment.}
\label{fig_analogy}
\end{figure}

All the previously mentioned convergence properties\footnote{
  Glossary: % Specific vocabulatory recap:
  To the \textbf{convergence laws} are associated the \textbf{convergence law constants} that are unknown and problem-dependent. It is expressed with respect to the \textbf{convergence parameter(s)}.
  The convergence laws' reciprocal function (with respect to the convergence parameter) is always bounded by the \textbf{query complexity}.
  The \textbf{free-parameters} are associated to the \textbf{equation-to-fit}.
  It is derived from the error model injected in the observable equation.
} and algorithms are detailled in \Cref{table_convergence}.
This list of algorithms is not exhaustive but highlights how the structure of some quantum routines can be explained by analogy with polynomial function approximations \Cref{fig_analogy}.
The convergence parameter of a quantum routine is analogous to the polynomial order, the gate (operation) error\footnote{
  The gate error \cite[section 4.5.3]{nielsen_quantum_2010} is the difference between the two quantum gates: $ \widehat{Er}(d) = \widehat{U} - \widehat{U}(d) $.
  For some routines, the three matrices possess the same eigenbasis, which implies that the error convergence only behaves as the difference of the eigenvalues: $ \lambda_{Er, i} \ketbra{\lambda_{i}} = |\lambda_{u, i} \ketbra{\lambda_{i}} - \lambda_{d, i} \ketbra{\lambda_{i}} | $.
  % The analogy with function polynomial approximation error is: $ \mathrm{Er}(d, x) = | \mathrm{f}(x) - \mathrm{poly}(d, x) | $.
} is analogous to the 'truncation' error $\varepsilon$, and the query complexity with the series rate of convergence.
For practical evaluation, the initial state vector is analogous to the value on which the polynomial is evaluated.
In both cases, evaluating a given state vector or a given value is necessary to study scalar convergence, rather than matrix gate error, or, by analogy, function error. 
To provide useful insights, this scalar convergence must be a semi-norm that effectively captures the gate error (referred to as the error semi-norm). % (in \Cref{table_convergence} is presented the total function error. % $ \varepsilon(d) = \int | \mathrm{f}(x) - \mathrm{poly}(d, x) | dx $

Most of the research works are interested in the asymptotic behavior of quantum routines or, study experimentally the convergence important problems \cite{childs_theory_2021, tarruell_quantum_2019, campbell_early_2022, stanisic_observing_2022}.
Protocols that enable the production of a sufficient quality gate approximation to continue a quantum algorithm while minimizing the convergence parameters (and thus the quantum circuit figure of merit) have only recently emerged.
Four articles concern \ac{pf} construction with guaranteed error.
Papers \cite{zhao_making_2023, ikeda_measuring_2024} focus on long-time Hamiltonian simulation and \cite{mehendale_estimating_2025} to Hamiltonian simulation as a subroutine for a \ac{qpe} algorithm.
The paper \cite{sennane_robustness_2026} also investigates free-parameters associated with \ac{qpe} including the Trotter number required to find the ground state eigenvalue.

\ac{neeqma} studies the convergence property of quantum routines with respect to their convergence parameters.
It allows the extraction of the value of the convergence law constants (unknown/free-parameters), which are hard to compute classically.
% Knowing these constants gives important information about the algorithm, query \& initial-state joint properties.
Particularly, it allows answering the question: what are the smallest parameters (shallowest circuit) that allow reaching a given error semi-norm in the gate approximation to construct the quantum circuit?
A first section describes the general \ac{neeqma} algorithm. % implementation.
The second part contains the results of \ac{neeqma}'s implementations for eigenvalues filtering by \ac{qsp} and also Hamiltonian Simulation by Trotterization.

\noindent\textbf{Summary of Contributions.} This work introduces \textit{Numerical Error Extraction by Quantum Measurement Algorithm} (\ac{neeqma}), a novel protocol designed to empirically determine convergence constants of quantum routines directly from \ac{qpu} measurements.
Observables with known statistical estimators that allow extraction of the convergence law constants are constructed from the gate error model; thus, \ac{neeqma} bypasses the need for costly classical computations of the numerical error.

\subsection{Research Questions and Objectives}
The primary objective of this paper is to address the problem of finding the convergence parameters numerical values without costly classical computation. %bridge the gap between theoretical asymptotic upper bounds and the practical performance of quantum routines.
Specifically, we work on the following research questions:
\begin{enumerate}
    \item How can we numerically extract problem-dependent convergence constants directly from a \ac{qpu} without relying on costly classical computation?
    \item How can we leverage these extracted constants to predict the minimum circuit depth (\textit{e.g.} Trotter number or polynomial degree) required to reach a target error semi-norm?
    \item Is \ac{neeqma} versatile on the different convergence-based quantum routine?
    Two different routines are experimentally tested: Hamiltonian simulation by \ac{pf} and \ac{qsp} to approximate a gate function. % To what extent are these empirical error models robust across different quantum routines like Hamiltonian simulation and \ac{qsp}?
\end{enumerate}

\section{\acl{neeqma}}

\begin{figure*}[tb]
\begin{center}
\resizebox{\linewidth}{!}{\includegraphics{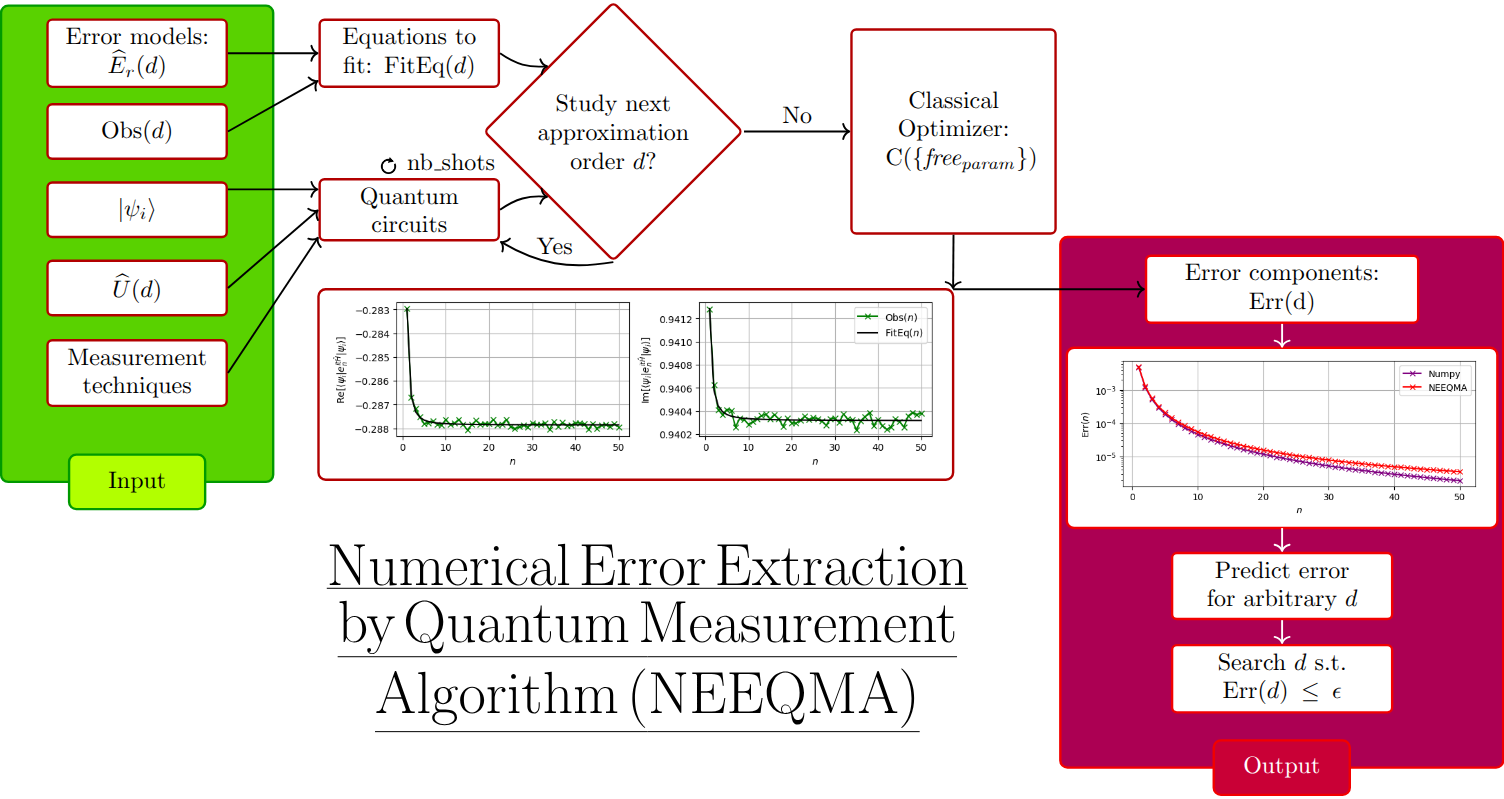}}
\end{center}
\caption{\ac{neeqma} Workflow.}
\label{workflow_neeqma}
\end{figure*}

The objective of \ac{neeqma} is to express a semi-norm of the numerical error with respect to the convergence parameter.
These values facilitate the retrieval of problem-dependent convergence law constants associated with the convergence law of the quantum routine.
It can be used to find the numerical values that guarantee that the quantum routine converges below a threshold.
Thus \ac{neeqma} does not generally “discover” convergence laws.
It performs parameter estimation for a computed parametric convergence model.
Unlike previous methods that rely on theoretical upper bounds, the \ac{neeqma} protocol introduced in this paper is a new approach to characterize routine-specific convergence.
While the workflow \Cref{workflow_neeqma} utilizes classical optimization as a sub-component, the overall methodology specifically the derivation of equations-to-fit from gate error-models is a novel contribution of this work.

The key to \ac{neeqma} is to select one (or a few) observable(s) that reflects the quantum routine convergence property when measured with respect to the convergence parameter $d$.
By injecting the error model into the observable's equation, it is possible to derive an equation-to-fit with free-parameters.
This equation-to-fit allows both for searching for the free-parameters that describe the \ac{qpu} outputs and for deducing the values of the convergence law constants from the free-parameters.
% The convergence law constants can then be injected into the error model.
One method to find candidate sets of free-parameters is to use classical optimization.
A choice of cost function to optimize is the difference between the observable measurement obtained thanks to the quantum circuit and the equation-to-fit at the same approximation order:

{\small
\begin{equation}
  \mathrm{C}(\{ \mathit{free_{param}} \}) 
  = \sum_{i = 1}^{d_{\mathit{max}}} | \mathrm{Obs}(i) - \mathrm{FitEq}(i, \{ \mathit{free_{param}} \}) |
  \label{eq_fit_obs}
\end{equation}}
with $d_{\mathit{max}}$ the largest value of $d$.

Inputs needed to run the \ac{neeqma} techniques include:
\begin{itemize}
  \item An equation-to-fit $ \mathrm{FitEq}(d) = \mathrm{Obs}[\widehat{U}(d) , \ket{\psi_{i}}] $ obtained from:
  \begin{itemize}
    \item The gate error model which is the gate dependence in the convergence parameter: $ \widehat{U}(d) = \widehat{U} + \widehat{Er}(d) $.
    \item Observables that can be experimentally measured on the circuit: $ \mathrm{Obs}[\widehat{U} , \ket{\psi_{i}}] $.
    It takes the measured quantum circuit as an input.
  \end{itemize}
  \item Quantum circuit execution:
  \begin{itemize}
    \item The initial state vector, from which starts the studied routine: $ \ket{\psi_{i}} $.
    \item The gate approximated by the routine with different values $d$ of the convergence parameter: $ \widehat{U}(d) $.
    \item The measurement circuit associated with $ \mathrm{Obs} $.
  \end{itemize}
\end{itemize}

\textit{Optional last step}
Once obtained, the convergence law constants are inserted into the error model to describe a semi-norm of the gate error on the initial state in relation to the convergence parameters.
As both the equation and constants of the convergence law are known, the curve can be extended to predict the error of a higher gate approximation order (\textit{i.e.} higher $d$).
The complete \ac{neeqma} workflow is detailed in \Cref{workflow_neeqma} and summarized in \Cref{table_neeqma}.

\begin{table}[h]
\centering
\begin{tabular}{|p{0.95\linewidth}|}
\hline
\textbf{Algorithm 1: NEEQMA Protocol} \\ \hline
\textbf{Inputs:} Initial state $\ket{\psi_{i}}$, gate model $\widehat{U}(d)$, Optional: target error semi-norm threshold $\epsilon$. \\
\textbf{1. Data Collection:} Measure observable $Obs(d)$ on QPU for small values of $d$ (e.g., $n \in [1, 13]$). \\
\textbf{2. Model Fitting:} Use a classical optimizer to fit $FitEq(d)$ to measurements and extract convergence law constants $\{\tilde{c}, \tilde{e}\}$. \\
\textbf{3. Extrapolation:} Compute the error model $\text{Err}(d)$ for any $d$ using the extracted convergence law constants. \\
\textbf{Output:} Plot a semi-norm of the error model $\text{Err}(d)$ for any $d$. \\
\textbf{Optional output:} Predict smallest $d$ such that $\text{Err}(d) \leq \epsilon$. \\ \hline
\end{tabular}
\label{table_neeqma}
\end{table}

\subsection{Comparison with Related Approaches}
While \ac{neeqma} shares technical similarities with \ac{zne}, such as the use of observables measured on different circuits that depend on a parameter $d$ \cite{giurgica-tiron_digital_2020}, its objective is fundamentally different.
\ac{zne} aims to extract \textbf{hardware-induced noise} by extrapolating to the zero-noise limit, whereas \ac{neeqma} extracts the convergence law constants of the algorithmic \textbf{numerical error} from the \ac{qpu}.

Compared to adaptive methods like Trotter24 \cite{ikeda_measuring_2024}, which iteratively adjust time-steps based on local error thresholds, \ac{neeqma} provides a global predictive model.
Once the constants are extracted, the optimal convergence parameter (\textit{e.g.} $n$ or $d$) can be determined for any target error $\epsilon$ without further quantum measurements.

\section{Experimental realization}
This section shows examples of \ac{neeqma}'s applications on two major quantum algorithms.
While the error models and observables can be reused, the problem instance was selected for illustrative purposes.
The cost associated with the observable sampling is not studied in these experiments.
This problem instance $ \widehat{H_{p}} = \sum_{i} \lambda_{i} \ketbra{\lambda_{i}} $ is in both cases the $LiH$ molecule with an inter-nucleus distance of $ 2 \mathring{\mathrm{A}} $.
The $LiH$ molecule at a separation of $ 2 \mathring{\mathrm{A}} $ was chosen as a benchmark system for illustrative purposes \cite{suzuki_generalized_1976}.
This geometry represents a stretched bond configuration where electron correlation effects are significant, providing a non-trivial test case for Hamiltonian dynamics.
The primary goal is not to discover new chemical properties of $LiH$, but to demonstrate that \ac{neeqma} can accurately extract convergence constants for typical molecular Hamiltonians where analytical bounds are too pessimistic.
How to obtain the corresponding Hamiltonian's Pauli decomposition is provided in the appendix.
The normalization factor $2|\lambda_{m}|$ is the norm of the largest eigenvalue and is computed classically.
In practical applications, the normalization factor is chosen to be a value exceeding the magnitude of the largest eigenvalue.
In both experiments, initial state vectors are the molecule \ac{hf} states: $ \ket{HF} = \ket{111000111000} $.

The paper appendix proposes: additional observables, details about the \ac{pf}, their associated error models, as well as experimentations.
The appendix also explains how to obtain the \ac{qsp} phase-angle series and provides experimental details.

\subsection{Hamiltonian Simulation by Trotterization}
\ac{pf} permits to construct the hamiltonian simulation $ e^{i t \widehat{H_{p}}} $ given the access to $ e^{i t \widehat{H_{j}}} $ where $ \widehat{H_{p}} = \sum_{j} \alpha_{j} \widehat{H_{j}} $.
The simplest \ac{pf} is the Lie-Trotter formula:
\begin{equation}
  e^{i t \widehat{H_{p}}} = \lim_{n \rightarrow \infty} e^{i t \widehat{H_{p}}}_{n} \text{, with: } e^{i t \widehat{H_{p}}}_{n} = \{ \prod_{j} e^{i \frac{t}{n} \alpha_{j} \widehat{H_{j}}} \}^{n}
\end{equation}
for which the gate error is mathematically derived using both the multiple applications of \ac{bch} formula and a Taylor series (the full computation is in the appendix \Cref{eq_error_model_trotter_lie_no_time}): % (truncated at the order one):
\begin{equation}
\begin{aligned}
  e^{i t \widehat{H_{p}}} & = e^{i \widehat{H_{p}}}_{n} + \frac{\widehat{Er_{1}}}{n} + \frac{\widehat{Er_{2}}}{n^{2}} + \mathcal{O}(\frac{1}{n^{3}}) \\
  & = e^{i t \widehat{H_{p}}}_{n} + \widehat{Er}(n)
  \label{eq_error_formula}
\end{aligned}
\end{equation}

\begin{figure}[tb]
\begin{center}
\resizebox{\linewidth}{!}{\includegraphics{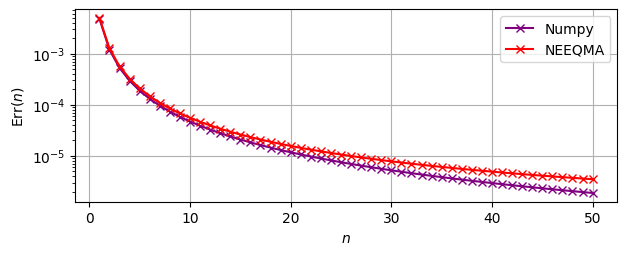}}
\end{center}
\caption{Error model with the convergence law constants obtained from the quantum circuit modelling and directly from matrix calculation (using numpy) with respect to the convergence parameter $n$ at the Hamiltonian simulation time $ t = 1 $.
Note that the increasing distance between the two curves is an artefact of the graph's log scale.}
\label{fig_hs_neeqma_results_t1}
\end{figure}

\begin{figure}[tb]
\begin{center}
\resizebox{\linewidth}{!}{\includegraphics{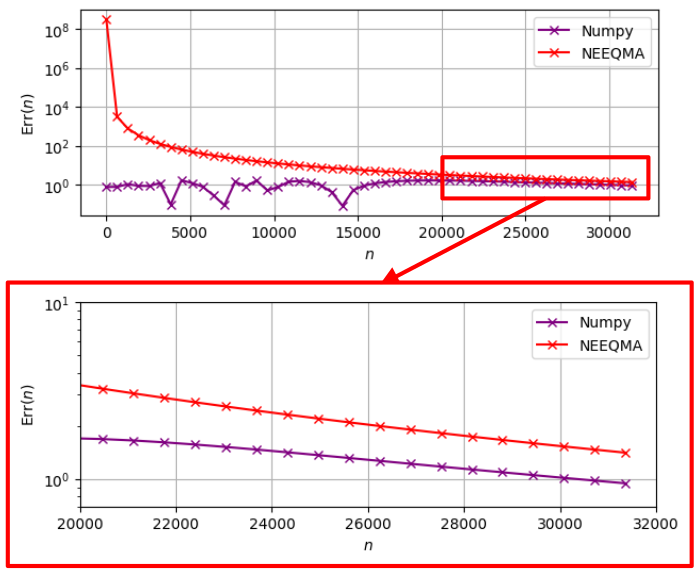}}
\end{center}
\caption{Error model with the convergence laws constants obtained from the quantum circuit modelling and directly from matrix calculation (numpy) with respect to the convergence parameter $n$ for an Hamiltonian simulation time $ t = 2 \pi \times 2^{10} $.}
\label{fig_hs_neeqma_results_4}
\end{figure}

The idea of this subsection is to model the evolution of the $LiH$ molecule's electrons under the molecule's potential starting from the \ac{hf} state.
This is done for a time $ t = 1 $ using the Lie-Trotter formula \Cref{fig_hs_neeqma_results_t1}.
The obtained convergence law constants are then used to extrapolate the error for a time $ t = 2 \pi \times 2^{10} \geq \frac{2 \pi}{\varepsilon} $ with $ \varepsilon = 10^{-3} $ the chemical accuracy \Cref{fig_hs_neeqma_results_4}.
This time corresponds to the maximum simulation time needed to construct a \ac{qpe} algorithm that decomposes the eigenstates with the error semi-norm (technically a component of the matrix error) $ | \bra{HF} \widehat{Er}(n) \ket{HF} | \leq \varepsilon $.

\begin{figure}[tb]
\begin{center}
\resizebox{\linewidth}{!}{\includegraphics{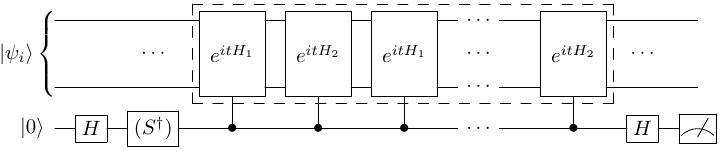}}
\end{center}
\caption{Trotter plus Hadamard-test quantum circuit.
Here: $ \widehat{H_{p}} = \widehat{H_{1}} + \widehat{H_{2}} $.
The dashed box delimits the Trotterization routine.
The $\widehat{S}^{\dag}$ gate is used to measure the imaginary part (but replaced by an identity gate when measuring the real part).}
\label{fig_qc_trotter}
\end{figure}

The chosen observables are the real and imaginary parts of the quantum circuit evaluated by the initial state, obtained by the Hadamard test \Cref{fig_qc_trotter}:
\begin{equation}
  \mathrm{Obs}(n) = 
%   \left\{
%     \begin{array}{ll}
        \left[ \mathrm{Re}[\bra{HF} e^{i t \widehat{H_{p}}}_{n} \ket{HF}] \textbf{, }
        \mathrm{Im}[\bra{HF} e^{i t \widehat{H_{p}}}_{n} \ket{HF}] \right]
%     \end{array}
%   \right.
\end{equation}
The selection of a suitable observable is guided by the target routine: for Hamiltonian simulation, the real and imaginary parts (measured via a Hadamard test) are natural candidates as they directly track the unitary evolution.
Reliability is tested by comparing the fitted parameters against using direct classical computation of the error.
It leads to the following equations-to-fit respectively:
\begin{equation}
  \begin{aligned}
  \mathrm{FitEq}(n) & = \left[ \mathrm{Re}[\bra{HF} e^{i t \widehat{H_{p}}}_{n} \ket{HF}] \textbf{, } \mathrm{Im}[\bra{HF} e^{i t \widehat{H_{p}}}_{n} \ket{HF}] \right] \\
  & \simeq \left[ cr + \frac{er_{1}}{n} + \frac{er_{2}}{n^{2}} \textbf{, } \qquad ci + \frac{ei_{1}}{n} + \frac{ei_{2}}{n^{2}} \right]
%   & \left\{
%     \begin{array}{ll}
%         \mathrm{Re}[\bra{HF} e^{i t \widehat{H_{p}}}_{n} \ket{HF}] & \simeq cr + \frac{er_{1}}{n} + \frac{er_{2}}{n^{2}} \\
%         \mathrm{Im}[\bra{HF} e^{i t \widehat{H_{p}}}_{n} \ket{HF}] & \simeq ci + \frac{ei_{1}}{n} + \frac{ei_{2}}{n^{2}}
%     \end{array}
%   \right.
  \end{aligned}
  \label{eq_fit}
\end{equation}
with the free-parameters (\textit{i.e.} convergence law constants) that are deduced: % for an arbitrary time:
\begin{equation}
\begin{aligned}
    cr & = \mathrm{Re}[\bra{HF} e^{i t \widehat{H_{p}}} \ket{HF}] \\
    ci & = \mathrm{Im}[\bra{HF} e^{i t \widehat{H_{p}}} \ket{HF}] \\
    er_{j} & = \mathrm{Re}[\bra{HF} \widehat{Er_{j}} \ket{HF}] \\
    ei_{j} & = \mathrm{Im}[\bra{HF} \widehat{Er_{j}} \ket{HF}] 
\end{aligned}
\label{eq_trotter_parameter_time_dep}
\end{equation}

\begin{figure}[tb]
\begin{center}
\resizebox{\linewidth}{!}{\includegraphics{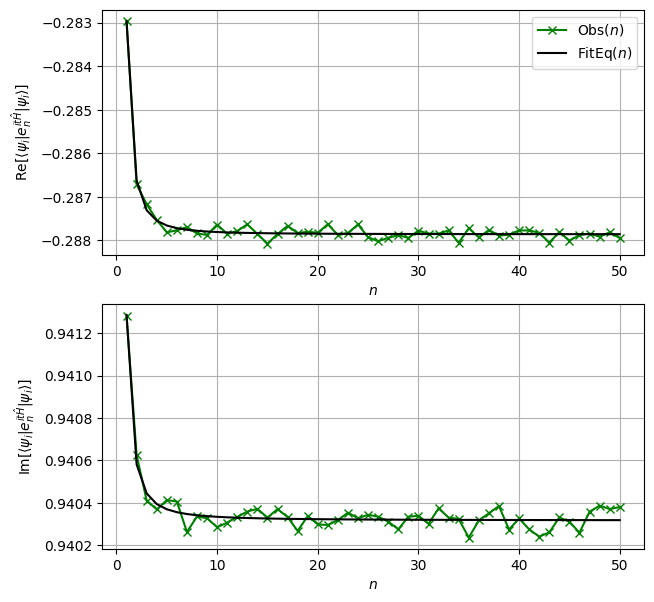}}
\end{center}
\caption{Observable measurements (real and imaginary part of the Trotterized Hamiltonian simulation) at different Trotter numbers $n$ with the associated error model, and with the free-parameters adjusted by classical optimization.
This curve was computed with a Hamiltonian simulation time $ t = 1 $.}
\label{fig_hs_fit_mosaique}
\end{figure}

% \textcolor{red}{The quantum circuits is similar to \Cref{fig_qc_trotter} and the curves fitted by classical optimization \Cref{fig_hs_fit_mosaique} leads to the free-parameters:
% \begin{equation}
% \begin{aligned}
%     \tilde{cr} & \simeq 2.88 \times 10^{-1} ; & \tilde{ci} & \simeq 9.4 \times 10^{-1} \\
%     \tilde{er}_{1} & \simeq 3.87 \times 10^{-5} ; & \tilde{ei}_{1} & \simeq 9.2 \times 10^{-5} \\
%     \tilde{er}_{2} & \simeq 4.84 \times 10^{-3} ; & \tilde{ei}_{2} & \simeq 8.8 \times 10^{-4}
% \end{aligned}
% \end{equation}
% The tilde is used to denote the value computed using \ac{neeqma}.}
The classical optimizer minimizes the cost function $C(\{free_{param}\})$ defined in \Cref{eq_fit_obs} by adjusting the free-parameters $\{free_{param}\} = \{c_r, c_i, e_{r1}, e_{i1}, e_{r2}, e_{i2}\}$ through the error models of \Cref{eq_fit} and defined in \Cref{eq_trotter_parameter_time_dep}.
This cost function allows for the error model to fit the experimental curves of the observables \Cref{fig_hs_fit_mosaique}.
These curves are obtained thanks to quantum circuits similar to the one of \Cref{fig_qc_trotter}.
Applying this procedure to the $LiH$ simulation yields the following numerical estimates (denoted with tildes):

\begin{table}[h]
\centering
\caption{Extracted Free Parameters for $LiH$ Trotterization ($t=1$)}
\begin{tabular}{ccc}
\hline
Constant & Real Part ($\tilde{r}$) & Imaginary Part ($\tilde{i}$) \\ \hline
$\tilde{c}$ & $2.88 \times 10^{-1}$ & $9.4 \times 10^{-1}$ \\
$\tilde{e}_1$ & $3.87 \times 10^{-5}$ & $9.2 \times 10^{-5}$ \\
$\tilde{e}_2$ & $4.84 \times 10^{-3}$ & $8.8 \times 10^{-4}$ \\ \hline
\end{tabular}
\label{tab:trotter_params}
\end{table}

These values represent the converged (during the classical optimization) convergence law constants that best describe the \ac{qpu} output for this specific problem instance.
 
The found model constants are injected into the error model \Cref{fig_hs_neeqma_results_t1}, which allows for expressing the error with respect to the convergence parameters:
\begin{equation}
\begin{aligned}
  \mathrm{Err}(n) & = | \bra{HF} \widehat{Er}(n) \ket{HF} | \\
  & \simeq | \frac{e_{1}}{n} + \frac{e_{2}}{n^{2}} |
\end{aligned}
\label{eq_error_model_trotter}
\end{equation}
with $ e_{j} = er_{j} + i ei_{j} $.

% More complex calculations allow us to show that the relation between the simulation time and the convergence parameters is bounded as shown by: (copier coller equation {eq_error_model_semi_bis}), it is possible to use the previous result to determine an error bound of our error norm for a short time in order to extrapolate for other times.
% The $\widehat{\Xi}$ matrices are expressed in terms of nested commutators of the problem Hamiltonian's summands.
As the relation between the simulation time and the convergence parameters is bounded as shown in appendix \Cref{eq_time_dep} (as derived in \Cref{eq_error_model_semi_bis}), it is possible to use the previous result to determine an error bound of our error semi-norm for a short time in order to extrapolate for other times:
% To this end, the time dependence, of the error model is :
\begin{equation}
%\begin{aligned}
    \frac{|e_{j}(t_{1})|}{t_{1}^{j + 1}} = \frac{|e_{j}(t_{2})|}{t_{2}^{j + 1}} % \\
%    \frac{ei_{j}(t_{1})}{t_{1}^{j + 1}} & = \frac{ei_{j}(t_{2})}{t_{2}^{j + 1}}
%\end{aligned}
\end{equation}

% \Cref{fig_hs_neeqma_results_4} curve analysis shows that the number of parameters needed to describe the curve increases with the Hamiltonian simulation time.
% It is particularly true for the first point of the curve as the ratio $ \mathrm{r}(n) = \frac{t^{j+1}}{n^{j}} $ is not, as for large $n$, dominated by $ j = 1$.
% Interestingly, it can be used to deduce the $e_{j}$ associated with large $j$ by progressively truncating the error model at higher Trotter numbers and increasing the simulation time.
% % There is no unknown to compute which power of $t^{2}$ is dominant on a given section of the convergence curve.

\begin{figure}[tb]
\begin{center}
\resizebox{\linewidth}{!}{\includegraphics{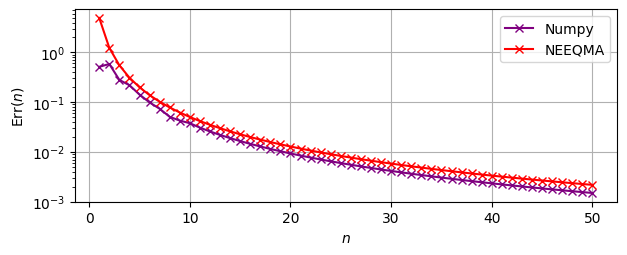}}
\end{center}
\caption{Error model with the convergence laws constants obtained from the quantum circuit modelling and directly from matrix calculation (numpy) with respect to the convergence parameter $n$ for an Hamiltonian simulation time $ t = 10 $.}
\label{fig_hs_neeqma_results_3}
\end{figure}

Looking at these results \Cref{fig_hs_neeqma_results_3} and \Cref{fig_hs_neeqma_results_4}, for low value of $ n < 20 000 $, the regime for which the convergence law truncation is accurate is not yet reached.
It leads to a difference between the truncated fitted model obtained by \ac{neeqma} and the exact computation.
To reach this regime at low $n$, one can use \ac{neeqma} to extract other convergence law constants associated with a higher-order truncation of the formula \Cref{eq_error_formula}.
It is difficult to explain why the lowest Trotter Error according to the NumPy simulation (purple curve) occurs near $ n = 4 000 $ and $ n = 14 000 $, given that one would typically expect the error to decrease as $n$ increases. This phenomenon might be explained by a specific addition of higher-order error terms that effectively cancel each other out at that specific point.

Once $n$ is high enough to reach the regime where the truncated convergence law \Cref{eq_error_formula} is correct: $n > 20 000$.
The \ac{neeqma} error semi-norm is as expected, an overbound of the computed error semi-norm by NumPy (that takes into account the higher order terms of \Cref{eq_error_formula}).

\subsection{Eigenvalue Filtering by \acl{qsp} \label{section_qsp}}
\begin{figure}[tb]
\begin{center}
\resizebox{\linewidth}{!}{\includegraphics{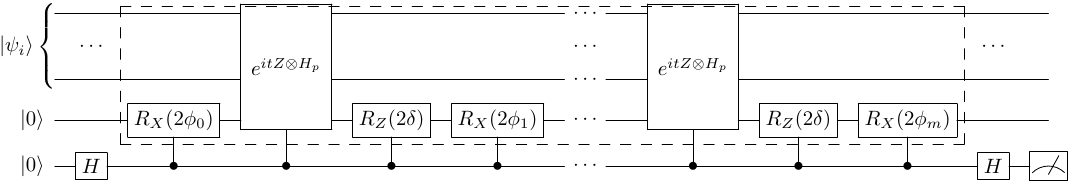}}
\end{center}
\caption{\ac{qsp} plus Hadamard-test quantum circuit.
The dashed box delimits the \ac{qsp} routine.}
\label{fig_qc_qsp}
\end{figure}

This subsection explores the \ac{isq} algorithm for searching eigenvalues of matrices \cite{dong_efficient_2021, ollive_quantum_2024}.
It is interesting because it requires only two qubits in addition to the one that encodes the problem.
It consists of doing the $\mathrm{Sign} \circ \cos $ function by \ac{qsp}. % in $W_{X}$ counvention.
Modifying a shift parameter on all the problem eigenvalues enables us to perform a binary search to separate the different eigenstates based on their associated eigenvalues.
% The following \ac{neeqma} implementation studies the splitting between the ground state and the first excited state (not degenerated with the ground state) of the LiH molecule. % with an inter-nucleus distance of $  $.
% In order to do it, the biger eigenvalues norm $|\lambda_{\mathit{max}}|$ used for the normalization factor $ \frac{\pi}{2|\lambda_{\mathit{max}}|} $ and the shift $ \delta = \frac{\pi}{2} + \frac{\pi}{2|\lambda_{\mathit{max}}|} \frac{|\lambda_{GS} + \lambda_{ES1}|}{2} $ are computed classically.
At the first iteration, the search space is split between positive and negative eigenvalues, and the problem Hamiltonian is not shifted:
\begin{equation}
\begin{aligned}
  \widehat{W_{Z}}[\widehat{H_{p}}] & = \exp( i t \widehat{H_{p}} \otimes \widehat{Z} ) \\
  & = 
  \begin{bmatrix}
      e^{i t \widehat{H_{p}}} & 0 \\
      0 & e^{- i t \widehat{H_{p}}}
  \end{bmatrix}
  \begin{matrix}
    \ket{HF} \ket{0} \\ % \ket{+} \\
    \ket{HF} \ket{1} % \ket{-}
  \end{matrix} \\
  & = \sum_{i} \ketbra{\lambda_{i}} \otimes
  \begin{bmatrix}
      e^{i t \lambda_{i}} & 0 \\
      0 & e^{- i t \lambda_{i}}
  \end{bmatrix}
  \begin{matrix}
    \ket{0} \\
    \ket{1}
  \end{matrix} \\
    & = \sum_{i} \ketbra{\lambda_{i}} \otimes
  \begin{bmatrix}
      \cos(t \lambda_{i}) & - i \sin(t \lambda_{i}) \\
      - i \sin(t \lambda_{i}) & \cos(t \lambda_{i})
  \end{bmatrix}
  \begin{matrix}
    \ket{+} \\
    \ket{-}
  \end{matrix} \\
  \text{where: } t & = \frac{\pi}{2 |\lambda_{\mathit{max}}|}
\end{aligned}
\label{eq_wz_hp}
\end{equation}
In \Cref{eq_wz_hp}, the state is defined on a composite Hilbert space $\mathcal{H}_{sys} \otimes \mathcal{H}_{anc}$, where $|HF\rangle \in \mathcal{H}_{sys}$ is the Hartree-Fock state of the LiH molecule (as detailed in Section III) and the second register corresponds to a single ancilla qubit.
The notation $|HF\rangle|0\rangle$ represents the tensor product $|HF\rangle \otimes |0\rangle$, where only the ancilla qubit is set to the computational zero state.
The following \ac{neeqma} implementation studies the second step of the binary search.
It splits the (normalized) eigenvalue space into two: the one under and the one above $ \Delta = -1/2 $.
The shifted signal operator is:
\begin{equation}
\begin{aligned}
  \widehat{W_{Z}}[\widehat{H_{p}} + \Delta \widehat{I}] & = \widehat{W_{Z}}[\widehat{H_{p}}] \widehat{R_{Z}}(\theta = -2 \delta) \\
  \text{where: } \delta & = \frac{\pi}{2} \Delta
\end{aligned}
\end{equation}
The initial state of the entire system is denoted as $|\psi_{init}\rangle = |HF\rangle \otimes |0\rangle$, where the index "init" refers to the starting state before the application of the QSP or Lie-Trotter sequence.
% In practical implementation, the shift is found by binary search.
The phase angles $\{ \phi_{d, i} \}$ and polynomial coefficients $ \mathrm{poly}(d, x) $ associated with the sign function polynomial approximation at different orders are obtained from \cite[pyqsp code]{chao_finding_2020, dong_efficient_2021, martyn_grand_2021}.
The associated quantum circuit \Cref{fig_qc_qsp}:
\begin{equation}
\begin{aligned}
\widehat{QSP}_{Z}[\mathrm{f}] & = \prod_{m = 1}^{m_{\phi}} \{ \widehat{S_{X}}(\phi_{m}) \widehat{W_{Z}} \} \widehat{S_{X}}(\phi_{0}) \\
& = \sum_{i} \ketbra{\lambda_{i}} \otimes
  \begin{bmatrix}
      \mathrm{F}(e^{i t \lambda_{i}}) & i \mathrm{G}(e^{i t \lambda_{i}}) \\
      i \mathrm{G}(e^{-i t \lambda_{i}}) & \mathrm{F}(e^{- i t \lambda_{i}})
  \end{bmatrix} \\
\end{aligned}
\end{equation}
with:
\begin{equation}
\begin{aligned}
\widehat{S_{X}}(\theta) & = e^{i \theta \widehat{X}} = \widehat{R_{X}}(-2 \theta)
% \widehat{S_{Z}}(\theta) & = e^{i \theta \widehat{Z}} = \widehat{R_{Z}}(-2 \theta)
\end{aligned}
\end{equation}

where $d$ is the order of the polynomial approximation of the function and $ \mathrm{F}, \mathrm{G} $ are real Laurent polynomials \cite[APPENDIX A]{martyn_grand_2021}.
It is the convergence parameter associated with the \ac{qsp} routine.
The chosen observable is a Hadamard test that measures the real part of the rest of the \ac{qsp} %post-select 
quantum circuit: % (joint probability of the real part when the first measurement indicates a projection into the block encodings' first component):
\begin{equation}
  \mathrm{Obs}(d) = 
%  \left\{
%    \begin{array}{ll}
        %\frac{
        \mathrm{Re}[\bra{\psi_{i}} \widehat{QSP^{d}_{\mathrm{sign}}}[\widehat{W_{Z}}] \ket{\psi_{i}}]%}{%} \\
        % \bra{0} \widehat{QSP^{d}_{\mathrm{sign}}}[\widehat{W_{Z}}] \ket{0}}
%    \end{array}
%  \right.
\end{equation}
The error model with respect to the approximation order $d$:
\begin{equation}
\begin{aligned}
%  \bra{0} \widehat{QSP^{d}_{\mathrm{sign}}} & [\widehat{W_{Z}}] \ket{0} \\
%  & = \sum_{x \in \{ \lambda \}} \mathrm{poly}_{\mathrm{sign}}(d, \cos(t x + \Delta)) \ketbra{x} \\
%  \begin{bmatrix}
%       \sum_{x \in \{ \lambda \}} \mathrm{poly}(d, \cos(\frac{\pi}{2} x)) \ket{x} \bra{x} % & . \\
%      . & .
%  \end{bmatrix} \\
%  \Rightarrow 
  \mathrm{FitEq} & (d, \{ \lambda \}, \{ \alpha_{\lambda} \}) \\
  & = \mathrm{Re}[\bra{\psi_{i}} \widehat{QSP^{d}_{\mathrm{sign}}}[\widehat{W_{Z}}] \ket{\psi_{i}}] \\
  & = \sum_{\lambda_{j} \in \{ \lambda \}} \alpha_{\lambda_{j}} \mathrm{Re}[\mathrm{F}(d, e^{i t \lambda_{j}})] \\
  & = \sum_{\lambda_{j} \in \{ \lambda \}} \alpha_{\lambda_{j}} \mathrm{poly}_{\mathrm{sign}}(d, \cos(t \lambda_{j} + \Delta))
\end{aligned}
\end{equation}
with: $ \alpha_{\lambda_{j}} = |\braket{ \lambda_{j} | HF }|^{2} \geq 0 $ and $ \sum_{\lambda_{j} \in \{ \lambda \}} \alpha_{\lambda_{j}} = 1 $.
If the polynomials implemented by \ac{qsp} are unknown, this cost function is computable thanks to a weighted sum of one-qubit-\ac{qsp} using the same series of phase angles.

\begin{figure}[tb]
\begin{center}
\resizebox{\linewidth}{!}{\includegraphics{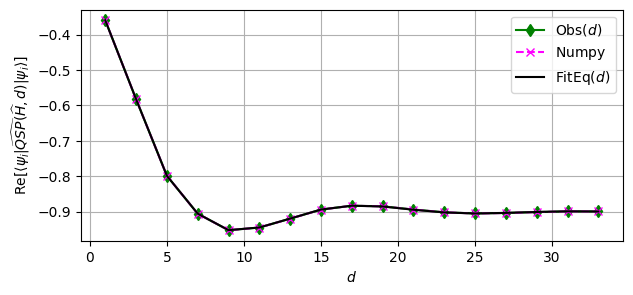}}
\end{center}
\caption{Observable measurements (real part of the \ac{qsp} of the Sign function) at different polynomial approximation orders $d$ with the associated error model, with the free-parameters adjusted by classical optimization.
Here $ k = 3 $ and the cost function value $ \mathrm{C} = 0.001929 $.
The numpy curve is computed using the $\mathrm{FitEq}$ function with the eigenvalues and $\alpha$ probabilities obtained by matrix diagonalization.}
\label{fig_qsp_fit}
\end{figure}

% with $\beta_{0}$ the probability to measure a \ac{qsp} output with the good matrix function, $\ket{0_{\perp}}$ contains all the states orthogonal to the $\ket{0}$ of the control register of the block encoding: $ 0 = \braket{0 | 0_{\perp}} $.
The free-parameters are the eigenvalues $\{ \lambda \}$ and the norm of the projection squared of the eigenstates on the initial state $\{ \alpha_{\lambda} \}$.
To fit this equation, we make the strong assumption that the convergence is primarily dictated by the $k$ eigenvalues associated with the eigenstates that have the most substantial overlap with the initial state, specifically those with the most significant overlap with the \ac{hf} state $\alpha$.
The error model constants are extracted from \Cref{fig_qsp_fit}:

\begin{table}[h]
\centering
\caption{Comparison of QSP Parameters: NEEQMA vs. Exact Diagonalization}
\begin{tabular}{ccccc}
\hline
$j$ & $\tilde{\alpha}_{\lambda_j}$ (NEEQMA) & $\alpha_{\lambda_j}$ & $\tilde{\lambda}_j$ (NEEQMA) & $\lambda_j$ \\ \hline
0 & 0.7597 & 0.4368 & $-1.259 \times 10^{-1}$ & $-1.270 \times 10^{-1}$ \\
1 & 0.2289 & 0.2616 & $-1.435 \times 10^{-1}$ & $-1.236 \times 10^{-1}$ \\
2 & 0.0112 & 0.1176 & $-1.207 \times 10^{-1}$ & $-1.331 \times 10^{-1}$ \\ \hline
\end{tabular}
\label{tab:qsp_params}
\end{table}

All the shifted eigenvalues are reported in \Cref{fig_analogy} (dashed vertical lines).

\begin{figure}[tb]
\begin{center}
\resizebox{\linewidth}{!}{\includegraphics{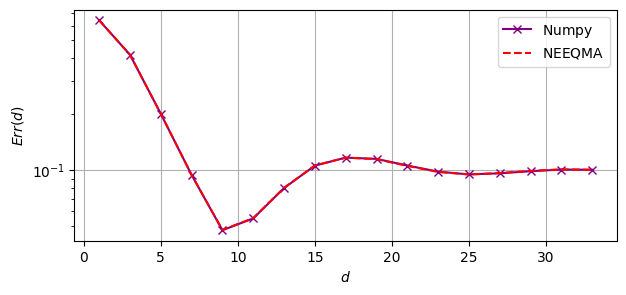}}
\end{center}
\caption{Error model with the convergence laws constants obtained from the quantum circuit modelling and directly from matrix diagonalization (numpy) with respect to the approximation order $d$.}
\label{fig_qsp_neeqma}
\end{figure}

Finally, the free-parameters are reinjected into the error model \Cref{fig_qsp_neeqma} to quantify the distance between the applied polynomial and the sign function:
\begin{equation}
\begin{aligned}
  & \mathrm{Err}(d) \simeq \\
  | \sum_{\lambda_{j} \in \{ \lambda \}_{m}} & \alpha_{\lambda_{j}} ( \mathrm{sign}(\cos(\frac{\pi}{2} \lambda_{j})) - \mathrm{poly}(d, \cos(\frac{\pi}{2} \lambda_{j})) ) |
\end{aligned}
\label{eq_error_model_qsp}
\end{equation}
As the polynomial can be generated for arbitrary $d$, this curve can be extended to study the algorithm convergence for larger $d$ values.
% The found model constants are injected into the error model, which allows for expressing the error with respect to the convergence parameters \Cref{fig_qsp_neeqma}.

% \textcolor{teal}{A key assumption in the \ac{qsp} implementation of \ac{neeqma} is that convergence is primarily driven by a subset of $m$ eigenvalues with the largest overlap ($\alpha_{\lambda_{j}}$) with the initial state. As shown in \Cref{fig_qsp_fit}, even with a small $m=3$, the optimizer successfully fits the observable curve with a cost function $C = 0.001929$, demonstrating robustness against the omission of low-weight eigenstates.}

% \textcolor{teal}{Potential model mismatch (e.g., higher-order Lie-Trotter terms) is mitigated by using the derived equation-to-fit from the \ac{bch} expansion, which accounts for terms up to $O(1/n^2)$. Statistical noise in the \ac{qpu} measurements is naturally handled by the classical optimizer, which minimizes the global residual error across all measured $d$ values.
% }

\section{Conclusion}

This paper presents a method for extracting information about the gate error constructed by quantum routines using carefully chosen observables that can be directly estimated on a gate-based \ac{qpu} capable of running the required gate sequence.
Often, it is known that a quantum routine converges, but not the speed of this convergence.
\ac{neeqma} addresses this issue by expressing carefully chosen observable's expectation values, which evolve depending on the gate error.
We also show that when a convergence law can be derived, it allows us to extract the constants of the error convergence law.
When injected into the error law, it yields a specific semi-norm of the gate error that allows extrapolations.
Two selected routines, Lie-Trotter Hamiltonian simulation and \ac{qsp}, are used to construct different quantum gates for illustrative purposes and are studied using the \ac{neeqma} protocol.
The derived equations are adaptable to other instances of these routines.

Even though we realized the experiments using Qiskit simulator with noise-free gate. %\ac{neeqma} usually accounts for \ac{qpu} errors (decoherence, gate noise) by treating the hardware output as the ground truth for the "actual" convergence on a specific device \cite{nielsen_quantum_2010}.
The classical optimizer fits the convergence law parameters obtained by measurement with statistical noise (see \Cref{fig_hs_fit_mosaique}), effectively capturing the combined effect of algorithmic 'truncation' error.
Note that this article focuses on extracting numerical error from quantum routines, but not on the statistical convergence of the observables estimators used to obtain it. For this reason, the experiments were conducted with a high number of shots. One can expect that observables converging rapidly are highly dependent on the instances studied, complicating this analysis; however, the different possible observables increase the likelihood that at least one will enable information extraction about numerical error with a reasonable number of shots.
It is also worth mentioning that for bigger Hamiltonians associated with more complex systems, the error is expected to converge more slowly with respect to the convergence parameter.
It implies that the measured states at each convergence parameter value are less similar than in a simpler system where convergence is reached more quickly.
Due to Shot-Noise-Limit (\textit{a.k.a.} Standard Quantum Limit), the more similar two states are, the harder it is to distinguish them statistically, which implies that the less efficient the error convergence (and so the bigger the system), the easier the \ac{neeqma} error extraction.

% To ensure that there is no sampling issue, soften the assumption about the \ac{qsp} initial state, or combine the effects of different routine approximation parameters are keep for further work to investigate.
Using a modified Hadamard test \ac{neeqma} allows to sample arbitrary components of the error model matrix that may be of interest to adjust classical computation of this matrices.
Knowing the convergence law constants is crucial for constructing quantum routines, but may also be of interest in classical computing.
% We hope that \ac{neeqma} extension for other quantum circuit generation routines will follow the one proposed in this paper.

\section{Funding}
This research work was supported in part by the French PEPR integrated project Etude de la PIle Quantique — EPIQ, (ANR-22-PETQ-0007), it was also supported in part by the French PEPR integrated project HQI (ANR-22-PNCQ-0002).

\section{Acknowledgement}
The authors thank Pierre-Emmanuel CLET for his correction in Appendix \ref{section_annexeb} calculus and for providing the induction argument that proves the $\Xi_{2}$ formula \Cref{eq_xi_proof}.

\bibliographystyle{ieeetr}
\bibliography{article_neeqma}

\appendix

\subsection{Hamiltonian Simulation Construction}
Two main families of strategies allow to construct a Hamiltonian Simulation gate (operation) on a quantum computer.
They differ from the input terms needed to assemble the final gate:
\begin{itemize}
    \item Strategies based on the problem Hamiltonian block encoding.
    Well adapted for \ac{ftqc}: it has the best asymptotic complexity but requires many qubits.
    These techniques approximate the Hamiltonian simulation by truncated series expansion:
    \begin{itemize}
        \item \ac{qsvt} (and \ac{qsp}) \cite{low_hamiltonian_2019, martyn_grand_2021}
        \item \ac{lcu} \cite{childs_hamiltonian_nodate}
        \subitem including Truncated \ac{ts} \cite{berry_simulating_2015}
    \end{itemize}
    \item Strategies based on the problem Hamiltonian, individual terms, or sum up the exact Hamiltonian simulation combination.
    The combination is achieved thanks to \ac{pf}, which derives from the Trotter formula.
    These \ac{pf} are classified by the number of parameters used to describe them:
    \begin{itemize}
        \item Lie-Trotter \ac{pf}
        \item Trotter-Suzuki \ac{pf}
%        \item \ac{pf}
        \item Multi-\ac{pf} (MPF): a combination a \acp{pf} in a \ac{lcu}.
        \item Dynamical-Multi-\ac{pf} (DMPF)
    \end{itemize}
\end{itemize}

\begin{figure}[tb]
\begin{center}
\resizebox{\linewidth}{!}{\includegraphics{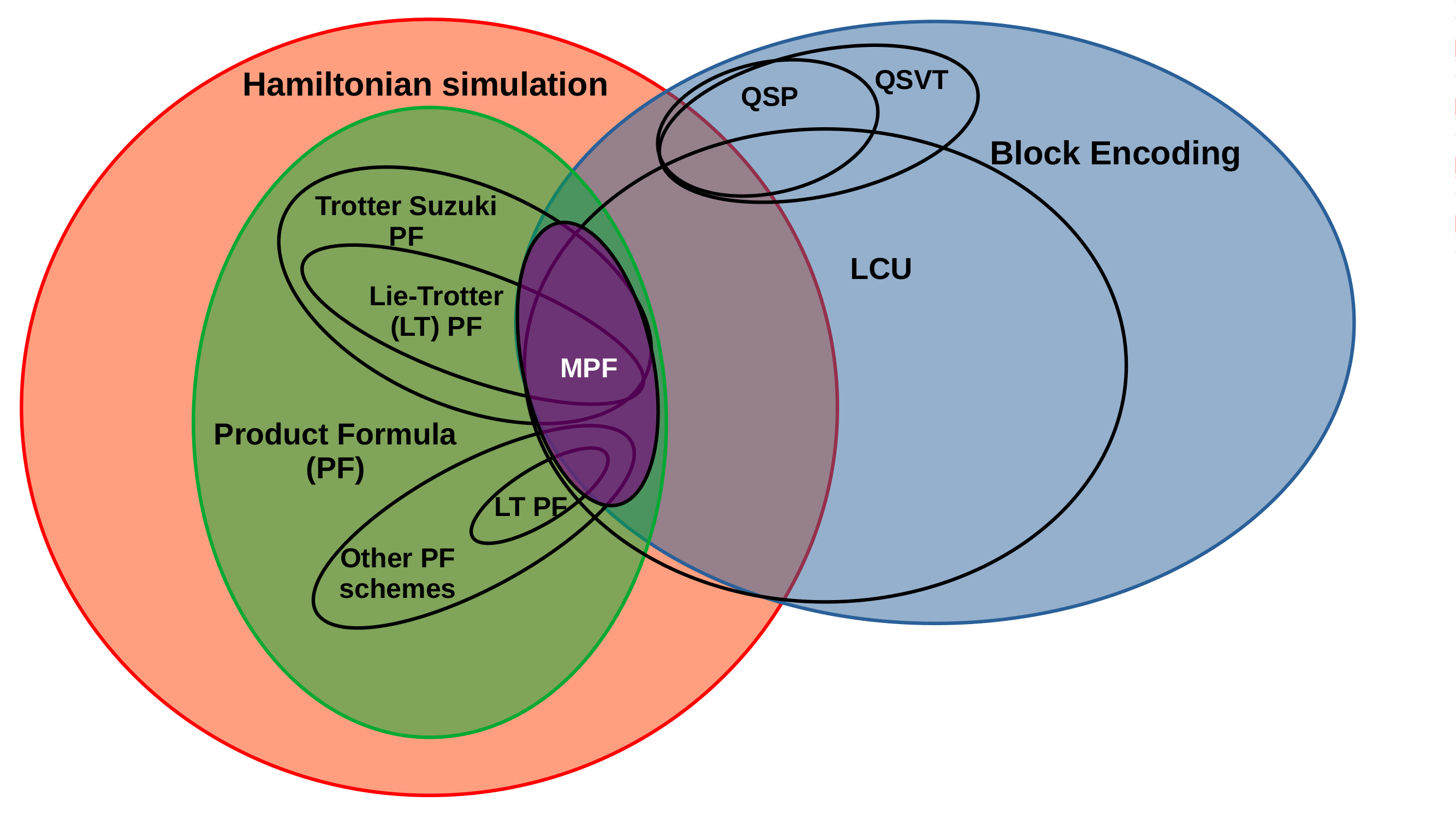}}
\end{center}
\caption{Schematic diagram of the different strategies to construct Hamiltonian Simulations.
When a strategy is included in a larger circle, it means this strategy is a specific instance of the larger circle strategy.}
\label{fig_product_formulas}
\end{figure}

\subsubsection{\acl{pf}}

$\widehat{S}(t) = e_{n,p}^{-i\widehat{H}t}$ will represent the \ac{pf} approximating the evolution of the Hamiltonian. The general formula for the \ac{pf} \cite[eq. 15]{childs_theory_2021}:

\begin{equation}
\begin{aligned}
    \widehat{S}(t):= \prod_{v=1}^\Upsilon \prod_{\gamma=1}^\Gamma e^{ta_{(v,\gamma)}\widehat{H}_{\pi_v(\gamma)}}
\end{aligned}
\end{equation}

'Where $\Upsilon$ is the number of stages of the formula, it relies on the Trotter number, when $p=1$, $\Upsilon = 1$, else $\Upsilon = 2 \times 5^{\frac{p}{2}-1}$. The permutation $\pi_v$ controls the ordering of operator summands within stage $v$ of the formula. The coefficients $a_{(v,\gamma)}$ are real numbers which rely on the Trotter order and the Trotter number' \cite[p. 13]{childs_theory_2021}. \\

\subsubsection{Multi product formula}

An extension to this formula introduces the \ac{mpf}\cite{low_well-conditioned_2019}:

\begin{equation}
\begin{aligned}
    % \widehat{S}(t) & = \sum_{i=1}^n c_i \widehat{S}(t) \\
    \widehat{S}(t) & = \sum_{i=1}^n c_i \prod_{v=1}^\Upsilon \prod_{\gamma = 1}^\Gamma e^{ta_{(i,v,\gamma)}\widehat{H}_{\pi_v(\gamma)}}
\end{aligned}
\end{equation}

The sum introduced by \ac{mpf} allows us to modify the Trotter number at each step. Then, using a linear combination, errors introduced by each term can be approximately canceled with a proper choice of $c_i$, reducing the Trotter error even further.

\subsubsection{Dynamical multi product formula}

Another extension of \ac{mpf} in \cite{zhuk_trotter_2024}, \ac{dmpf}:

\begin{equation}
\begin{aligned}
    % \widehat{S}(t) & = \sum_{i=1}^n c_i(t) \widehat{S}(t) \\
    \widehat{S}(t) & = \sum_{i=1}^n c_i(t) \prod_{v=1}^\Upsilon \prod_{\gamma=1}^\Gamma e^{ta_{(i,v,\gamma)}\widehat{H}_{\pi_v(\gamma)}}
\end{aligned}
\end{equation}

In this formula, the Trotter number changes with each term of the sum, and depending on the considered interval of time. It has been demonstrated in \cite{zhuk_trotter_2024} that it achieves better approximations than \ac{mpf}. To make dynamic multi-product formulas resilient to uncertainty from algorithmic errors, sampling, and hardware noise, they used a minimax estimation method (Minimax \ac{mpf}) \cite[Section IV]{zhuk_trotter_2024}.

\subsection{Trotter Error Model Derivation \label{section_annexeb}}
Error models associated with arbitrary \ac{pf} are in general difficult to compute.
Lie-Trotter \ac{pf}'s error model is a specific case that admits an analytic truncated approximation.
This section details this computation using this paper's notations:
\begin{itemize}
    \item The \ac{bch}'s formula \cite{achilles_early_2012}:
    \begin{equation*}
        e^{\widehat{C}} = e^{\widehat{A}} e^{\widehat{B}}
    \end{equation*}
    with $ \widehat{C}(\widehat{A}, \widehat{B}) = \widehat{A} + \widehat{B} + \frac{[\widehat{A}, \widehat{B}]}{2} + \frac{[\widehat{A} [\widehat{A}, \widehat{B}]] - [\widehat{B}, [\widehat{A}, \widehat{B}]]}{12} + \dots $.
    Note that $ \widehat{C}(\widehat{A}, \widehat{B}) = \widehat{A} + \widehat{B} $ when $\widehat{A}$ and $\widehat{B}$ commute.
    \item Recursive reduction of the Hamiltonian's Hamiltonian simulation to \ac{bch}:
    \begin{equation*}
    \begin{aligned}
        \widehat{H} & = \sum_{j = 0}^{N} \widehat{H_{j}} \\
        \Rightarrow e^{\widehat{C}} & = \prod_{j = 0}^{N} e^{t \widehat{H_{j}}} = e^{t \widehat{H_{0}}} \prod_{j = 1}^{N} e^{t \widehat{H_{j}}}
    \end{aligned}
    \end{equation*}
    with:
    \begin{equation*}
    \begin{aligned}
        \widehat{C} & = \widehat{C}(t \widehat{H_{0}}, \widehat{C}(t \widehat{H_{1}}, \widehat{C}(t \widehat{H_{2}}, \widehat{C}(\dots, \widehat{C}(t \widehat{H_{N-1}}, t \widehat{H_{N}}) \dots )))) \\
        & = t \widehat{H} + t^{2} \widehat{\Xi_{1}} + t^{3} \widehat{\Xi_{2}} + \dots
    \end{aligned}
    \end{equation*}
    with:
    \begin{equation*}
    \begin{aligned}
        \widehat{\Xi_{1}} & = \frac{1}{2} \sum_{j < k} [\widehat{H_{j}}, \widehat{H_{k}}] \\
        \widehat{\Xi_{2}} & = \frac{1}{12} ( \sum_{j} \sum_{k} [\widehat{H_{j}}, [\widehat{H_{j}}, \widehat{H_{k}}]] \\
        & \qquad \quad + 2 \sum_{j < k < l} ( [\widehat{H_{j}}, [\widehat{H_{k}}, \widehat{H_{l}}]] + [\widehat{H_{l}}, [\widehat{H_{k}}, \widehat{H_{j}}]] ))
    \end{aligned}
    \end{equation*}
    \subitem It is based on $ e^{\widehat{A}} e^{\widehat{B}} e^{\widehat{D}} = e^{\widehat{X}} e^{\widehat{D}} = e^{\widehat{Y}} $ with $ \widehat{X} = \widehat{C}(\widehat{A}, \widehat{B}) $ and $ \widehat{Y} = \widehat{C}(\widehat{X}, \widehat{D}) $.
    \item Trotter repetitions:
    \begin{equation*}
        (\prod_{j = 0}^{N} e^{\frac{t}{n} \widehat{H_{j}}})^{n} = (e^{\widehat{C}(n)})^{n}
    \end{equation*}
    with $ \widehat{C}(n) = \frac{t}{n} \widehat{H} + (\frac{t}{n})^{2} \widehat{\Xi_{1}} + (\frac{t}{n})^{3} \widehat{\Xi_{2}} + \dots $ and
    \begin{equation*}
    \begin{aligned}
    n \widehat{C}(n) & = t \widehat{H} + \frac{t^{2}}{n} \widehat{\Xi_{1}} + \frac{t^{3}}{n^{2}} \widehat{\Xi_{2}} + \dots \\
    & = t \widehat{H} + \sum_{m = 1}^{\infty} \frac{t^{m+1}}{n^{m}} \widehat{\Xi_{m}} \\
    & = t \widehat{H} + \widehat{\Xi}(n, t)
    \end{aligned}
    \end{equation*}
    It is enough to understand that $ \lim_{n \rightarrow \infty} \widehat{\Xi}(n, t) = 0 $ which implies:
    \begin{equation*}
      e^{i t \widehat{H}} = \lim_{n \rightarrow \infty} \{ \prod_{j} e^{i \frac{t}{n} \widehat{H_{j}}} \}^{n}
    \end{equation*}
    \item Using Taylor series: % If $n$ is large enough, $ \widehat{\Xi}(n, t)^{2} << \widehat{\Xi}(n, t) $ which implies:
    \begin{equation}
    \begin{aligned}
        e^{t \widehat{H} + \widehat{\Xi(n, t)}} & = \sum_{k = 0}^{\infty} \frac{(t \widehat{H} + \widehat{\Xi}(n, t))^{k}}{k!} \\
        & = \sum_{k = 0}^{\infty} \frac{(t \widehat{H})^{k}}{k!} + \sum_{k = 1}^{\infty} \frac{1}{k!} \sum_{w \in \mathcal{W}_{k}} \prod_{w_{l} \in w} w_{l} \\
        & = e^{t \widehat{H}} + \sum_{m = 1}^{\infty} \frac{1}{n^{m}} \widehat{Er}_{m} \\
        & \simeq e^{t \widehat{H}} + \widehat{Er}(n, t)
    \end{aligned}
    \label{eq_error_model_trotter_lie_no_time}
    \end{equation}
    with:
    \begin{equation*}
        \mathcal{W}_{k} = \{ t \widehat{H}, \widehat{\Xi}(n, t) \}^{k} \text{\textbackslash} \{ t \widehat{H} \}^{k}
    \end{equation*}
    when $ m \leq 2 $, $ \widehat{Er_{m}} $ is written explicitly as:
    \begin{equation*}
    \begin{aligned}
        \widehat{Er}_{1} & = \widehat{E}_{1, 1} \\
        \widehat{Er}_{2} & = \widehat{E}_{2, 1} + \widehat{E}_{1, 2} \\
        \widehat{E}_{m, 1} & = \sum_{k = 1}^{\infty} \frac{t^{k + m} \sum_{l = 0}^{k - 1} \widehat{H}^{l} \widehat{\Xi}_{m} \widehat{H}^{k - l - 1}}{k!} \\
        \widehat{E}_{m, 2} & = \sum_{k = 2}^{\infty} \frac{t^{k + 2 m}}{k!} \sum_{l + j \leq k-2} \widehat{H}^{j} \widehat{\Xi}_{m} \widehat{H}^{l} \widehat{\Xi}_{m} \widehat{H}^{k - l - j - 2}
        \label{eq_time_dep}
    \end{aligned}
    \end{equation*}
    \item note that (with $||.||$ a matrix sub-multiplicative norm):
    \begin{equation}
    \begin{aligned}
        ||\widehat{Er}_{1}|| & \leq t^{2} ||\widehat{\Xi}_{1}|| e^{t ||\widehat{H}||} \\
        ||\widehat{Er}_{2}|| & \leq t^{3}(||\widehat{\Xi}_{2}|| + t \frac{||\widehat{\Xi}_{1}||^{2}}{2}) e^{t ||\widehat{H}||}
    \end{aligned}
    \label{eq_error_model_trotter_lie}
    \end{equation}
    Moreover, for a matrix semi-norm (not necessarily sub-multiplicative) defined as the expectation value $ |\bra{\psi} . \ket{\psi}| $, it is possible to write:
    \begin{equation}
    \begin{aligned}
        |\bra{\psi} \widehat{Er}_{1} \ket{\psi}| & \leq t^{2} |\bra{\psi}  \widehat{\Xi}_{1} \ket{\psi}| + \mathrm{\Delta_{1}}(t) \\
        | \bra{\psi} \widehat{Er}_{2} \ket{\psi} | & \leq t^{3}| \bra{\psi} \widehat{\Xi}_{2} \ket{\psi}| + \mathrm{\Delta_{2}}(t) \\
    \end{aligned}
    \label{eq_error_model_semi}
    \end{equation}
    with:
    \begin{equation}
    \begin{aligned}
      \mathrm{\Delta_{1}}(t) & = || \sum_{k = 2}^{\infty} \frac{t^{k + 1}}{k!} \sum_{l = 0}^{k - 1} \widehat{H}^{l} \widehat{\Xi}_{1} \widehat{H}^{k - l - 1} || \\
      & = \mathrm{O}(t^{3}) \\
      \mathrm{\Delta_{2}}(t) & = || \sum_{k = 2}^{\infty} \frac{t^{k + 2}}{k!} \sum_{l = 0}^{k - 1} \widehat{H}^{l} \widehat{\Xi}_{2} \widehat{H}^{k - l - 1} || \\
      & \qquad + || \sum_{k = 2}^{\infty} \frac{t^{k + 2}}{k!} \sum_{l + j \leq k-2} \widehat{H}^{j} \widehat{\Xi}_{1} \widehat{H}^{l} \widehat{\Xi}_{1} \widehat{H}^{k - l - j - 2} || \\
      & = \mathrm{O}(t^{4}) \\
    \end{aligned}
    \end{equation}
    similarly:
    \begin{equation}
    \begin{aligned}
      | |\bra{\psi} \widehat{Er}_{1} \ket{\psi}| - t^{2} |\bra{\psi}  \widehat{\Xi}_{1} \ket{\psi}| | & = \mathcal{O}(t^{3}) \\
      | | \bra{\psi} \widehat{Er}_{2} \ket{\psi} |
      - t^{3} | \bra{\psi} \widehat{\Xi}_{2} \ket{\psi}| | & = \mathcal{O}(t^{4}) \\
      \label{eq_error_model_semi_bis}
    \end{aligned}
    \end{equation}
    % And $\widehat{Er}_{m}$ comes from:
    % For short simulation times $ t << 1 \Rightarrow \widehat{\Xi}(n, t)^{2} << \widehat{\Xi}(n, t) $, it simplify to:
    % \begin{equation}
    % \begin{aligned}
    %     e^{t \widehat{H} + \widehat{\Xi(n, t)}} & = \sum_{k = 0}^{\infty} \frac{(t \widehat{H})^{k}}{k!} + \sum_{k = 1}^{\infty} \frac{\sum_{l = 0}^{k - 1} (t \widehat{H})^{l} \widehat{\Xi}(n, t) (t \widehat{H})^{k - l - 1}}{k!} \\
    %     & \qquad \qquad + \mathcal{O}(\widehat{\Xi}(n, t)^{2}) \\
    %     & = e^{t \widehat{H}} + \sum_{m = 1}^{\infty} \frac{t^{m+1}}{n^{m}} \widehat{Er_{m}} + \mathcal{O}(\widehat{\Xi}(n, t)^{2}) \\
    %     & \simeq e^{t \widehat{H}} + \widehat{Er}(n, t) \\
    %     \text{with: } \widehat{Er}_{m} & = \sum_{k = 1}^{\infty} \frac{\sum_{l = 0}^{k - 1} (t \widehat{H})^{l} \widehat{\Xi}_{m} (t \widehat{H})^{k - l - 1}}{k!}
    % \end{aligned}
    % \end{equation}
    \item The complete proof of $\widehat{\Xi_{2}}$ formula is detailed in \Cref{eq_xi_proof}. %the associated file:
    \item $\widehat{Er}_{1}$ is also computed in \cite[eq.(1)]{lloyd_universal_1996} and $\widehat{\Xi}$ in \cite[eq.A18]{mehendale_estimating_2025}.
    \item Note that \cite[eq.(145)]{childs_theory_2021} propose a bound for $||\widehat{Er}(n=1, t)||$.
    % \begin{verbatim}
    %     'calcul_Xi2.pdf'.
    % \end{verbatim}
    \item An alternative proof of the Lie-Trotter convergence is detailed in \cite[section 4.7.2]{nielsen_quantum_2010}. It also proposes to use the Zassenhaus formula to prove that the Trotter-Suzuki formula with the Trotter number $n$ and Trotter order equal to $2$ has faster convergence than the Lie-Trotter formula (Trotter order equal to $1$).
\end{itemize}

\subsection{Other Observables to Measure the Error Convergence \label{section_fidelity}}
This section details the derivation of equations-to-fit associated with other observables to run \ac{neeqma} to construct the Hamiltonian Simulation.
% This section contains formulas to measure expectation values of different observables and their theoretical approximations to be easily fitted with the optimizer.
% Different \ac{ld} order of the Trotter error formula with respect to the \ac{bch} formula were considered.

These equations are derived for the Hamiltonian simulation error model truncated at order two: % one or two:
% \begin{itemize}
% \item Error model:
\begin{equation*}
\begin{aligned}
    & e^{-i t \widehat{H}} = e_{n}^{-i t \widehat{H}} - \frac{\widehat{Er_{1}}}{n} - \frac{\widehat{Er_{2}}}{n^{2}} + O(\frac{1}{n^{3}}) \\
% \end{equation*}
% % \item Order 1 error model:
% % \begin{equation*}
% %     e_{n}^{-i t \widehat{H}} \approx e^{-i t \widehat{H}} - \frac{\widehat{Er_{1}}}{n}
% % \end{equation*}
% \item Order 2 error model:
% \begin{equation*}
    \Leftrightarrow & e_{n}^{-i t \widehat{H}} \approx e^{-i t \widehat{H}} + \frac{\widehat{Er_{1}}}{n} + \frac{\widehat{Er_{2}}}{n^{2}}
\end{aligned}
\end{equation*}
% The second order is the one we used, it allowed us to have really good fit compared to the first order model.
% \end{itemize}

\begin{figure}[tb]
    \begin{center}
    \resizebox{\linewidth}{!}{\includegraphics{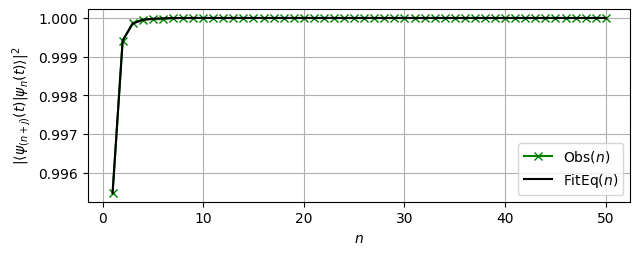}}
    \end{center}
    \caption{Observable measurement (fidelity between the Trotterized Hamiltonian simulation at trotter number $ n $ and $ (n+j) $, $ j = 1 $) at different Trotter numbers $ n $ with the associated error model, with the free-parameters adjusted by classical optimization.
    This curve was computed at Hamiltonian simulation's time $ t = 1 $.}
    \label{fig_qc_hs_fit_fidelity.png}
\end{figure}

Equations-to-fit derived using the error model:
\begin{itemize}
    \item The fidelity error\footnote{This technique possesses similarities with the technique used in the error-mitigation \ac{zne} method \cite[eq. 1]{giurgica-tiron_digital_2020}.
    It exploits the identity $ \widehat{U}^\dagger \widehat{U} = \widehat{I} $ to detect quantum noise arising during gate execution.
    Here, two different approximation orders of the Hamiltonian simulation are compared to determine if they differ.
    This observable is used in \cite{ikeda_measuring_2024}.} between a Trotterization of Trotter number $ n $, and a Trotterization of Trotter number $ n + j $ with $ j \in \mathbb{N} $:
    {\footnotesize \begin{equation*}
    \begin{aligned}
        & |\braket{ \psi_{n + j}(t) | \psi_{n}(t) }|^2 \\
        & \approx |\braket{ \psi_{0} | (e^{-i t \widehat{H}} + \frac{\widehat{E}r_{1}}{n + j} + \frac{\widehat{E}r_{2}}{(n + j)^2})^\dagger (e^{-i t \widehat{H}} + \frac{\widehat{Er}_{1}}{n} + \frac{\widehat{Er}_{2}}{n^{2}}) | \psi_{0} }|^2 \\
        & = |1 + \langle \psi_{0} | \frac{e^{-i t \widehat{H}} \widehat{E}r_{1}}{n} | \psi_0 \rangle + \langle \psi_0 | \frac{e^{-i t \widehat{H}}\widehat{E}r_{2}}{n^2} | \psi_0 \rangle + \langle \psi_0 | \frac{\widehat{E}r_{1}^\dagger e^{-i t \widehat{H}}}{n+j} | \psi_0 \rangle\\ & + \langle \psi_0 | \frac{\widehat{E}r_{1}^\dagger \widehat{E}r_{1}}{n^2+nj}| \psi_0 \rangle + \langle \psi_0 | \frac{\widehat{E}r_{1}^\dagger \widehat{E}r_{2}}{n^3+n^2j} | \psi_0 \rangle + \langle \psi_0 | \frac{\widehat{E}r_{2}^\dagger e^{-i t \widehat{H}}}{(n+j)^2} | \psi_0 \rangle \\ & + \langle \psi_0 | \frac{\widehat{E}r_{2}^\dagger \widehat{E}r_{1}}{(n+j)^2n} | \psi_0 \rangle + \langle \psi_0 | \frac{\widehat{E}r_{2}^\dagger \widehat{E}r_{2}}{(n+j)^2n^2} | \psi_0 \rangle |^2\\
        & = |1 + \frac{\mathit{cst_{1}}}{n} + \frac{\mathit{cst_{2}}}{n^2} + \frac{\mathit{cst_{1}}^*}{n+j} + \frac{\mathit{cst_{3}}}{n^2+nj} + \frac{\mathit{cst_{4}}}{n^3+n^2j}\\ & + \frac{\mathit{cst_{2}}^*}{(n+j)^2} + \frac{\mathit{cst_{4}}^{*}}{(n + j)^2n} + \frac{\mathit{cst_{5}}}{(n+j)^2n^2}|^2
    \end{aligned}
    \end{equation*} }
    with $ \mathit{cst_{1}}, \mathit{cst_{2}}, \mathit{cst_{4}} \in \mathbb{C} \: , \: \mathit{cst_{3}}, \mathit{cst_{5}} \in \mathbb{R} $:
    \begin{equation*}
    \begin{aligned}
        & \mathit{cst_{1}} = \langle \psi_0 | e^{-i t \widehat{H}}\widehat{E}r_{1} | \psi_0 \rangle\\
        & \mathit{cst_{2}} = \langle \psi_0 | e^{-i t \widehat{H}}\widehat{E}r_{2} | \psi_0 \rangle\\
        & \mathit{cst_{3}} = \langle \psi_0 | \widehat{E}r_{1}^\dagger \widehat{E}r_{1}| \psi_0 \rangle = |\widehat{E}r_{1}|^2 \in \mathbb{R}\\
        & \mathit{cst_{4}} = \langle \psi_0 | \widehat{E}r_{1}^\dagger \widehat{E}r_{2}| \psi_0 \rangle\\
        & \mathit{cst_{5}} = \langle \psi_0 | \widehat{E}r_{2}^\dagger \widehat{E}r_{2}| \psi_0 \rangle \in \mathbb{R}\\
    \end{aligned}
    \end{equation*}
    \Cref{fig_qc_hs_fit_fidelity.png} derives from \ac{neeqma} application with this observable.
    \item Measuring the Hamiltonian energy directly by protective measurement (this observable is particularly long to compute due to the need for the expectation value of many measurement circuits):
    % \subitem Order 1:
    % \begin{equation}
    % \begin{aligned}
    %     & \langle \psi_0 | (e_n^{-i t \widehat{H}})^\dagger \widehat{H} (e_n^{-i t \widehat{H}}) | \psi_0 \rangle \\
    %     & \approx \langle \psi_0 | (e^{-i t \widehat{H}})^\dagger \widehat{H} (e^{-i t \widehat{H}}) | \psi_0 \rangle + \langle \psi_0 | (e^{-i t \widehat{H}})^\dagger \widehat{H} \frac{(-1)}{n} \widehat{E}r_{1} | \psi_0 \rangle\\ & + \langle \psi_0 | \frac{(-1)}{n} \widehat{E}r_{1}^\dagger \widehat{H} e^{-i t \widehat{H}} | \psi_0 \rangle\\ & + \langle \psi_0 | \frac{(-1)}{n} \widehat{E}r_{1}^\dagger \widehat{H} \frac{(-1)}{n} \widehat{E}r_{1} | \psi_0 \rangle\\
    %     & = \mathit{cst_{1}} + \frac{\mathit{cst_{2}}}{n} + \frac{\mathit{cst_{2}}^*}{n} + \frac{\mathit{cst_{3}}}{n^2}\\
    %     & = \mathit{cst_{1}} + \frac{2}{n}\mathrm{Re}[\mathit{cst_{2}}] + \frac{\mathit{cst_{3}}}{n^2}\\
    %     & \mathit{cst_{1}}, \mathit{cst_{3}} \in \mathbb{R}, \mathit{cst_{2}} \in \mathbb{C}\\
    %     & \mathit{cst_{1}} = \langle \psi_0 | \widehat{H} | \psi_0 \rangle\\
    %     & \mathit{cst_{2}} = -\langle \psi_0 | (e^{-i t \widehat{H}})^\dagger \widehat{H} \widehat{E}r_{1} | \psi_0 \rangle\\
    %     & \mathit{cst_{3}} = (\langle \psi_0 |  \widehat{E}r_{1}^\dagger \widehat{H} \widehat{E}r_{1} | \psi_0 \rangle)
    % \end{aligned}
    % \end{equation}
    % \subitem Order 2:
    {\scriptsize \begin{equation*}
    \begin{aligned}
        & \langle \psi_0 | (e_n^{-i t \widehat{H}})^\dagger \widehat{H} (e_n^{-i t \widehat{H}}) | \psi_0 \rangle \\
        & \approx \langle \psi_0 | (e^{-i t \widehat{H}} + \frac{\widehat{E}r_{1}}{n} + \frac{\widehat{E}r_{2}}{n^2})^\dagger \widehat{H} (e^{-i t \widehat{H}} + \frac{\widehat{E}r_{1}}{n} + \frac{\widehat{E}r_{2}}{n^2}) | \psi_0 \rangle \\
        & = \braket{ \psi_0 | e^{-i t \widehat{H}}\widehat{H}e^{-i t \widehat{H}} | \psi_0 }  + \braket{ \psi_0 | \frac{e^{-i t \widehat{H}}\widehat{H}\widehat{E}r_{1}}{n} | \psi_0 } \\
        & + \braket{ \psi_0 | \frac{e^{-i t \widehat{H}} \widehat{H} \widehat{E}r_{2}}{n^2} | \psi_0 } + \langle \psi_0 | \frac{\widehat{E}r_{1}^\dagger \widehat{H} e^{-i t \widehat{H}}}{n} | \psi_0 \rangle \\
        & + \langle \psi_0 | \frac{\widehat{E}r_{1}^\dagger \widehat{H} \widehat{E}r_{1}}{n^2}| \psi_0 \rangle + \langle \psi_0 | \frac{\widehat{E}r_{1}^\dagger \widehat{H} \widehat{E}r_{2}}{n^3} | \psi_0 \rangle \\
        & + \langle \psi_0 | \frac{\widehat{E}r_{2}^\dagger \widehat{H} e^{-i t \widehat{H}}}{n^2} | \psi_0 \rangle + \langle \psi_0 | \frac{\widehat{E}r_{2}^\dagger \widehat{H} \widehat{E}r_{1}}{n^3} | \psi_0 \rangle + \langle \psi_0 | \frac{\widehat{E}r_{2}^\dagger \widehat{H} \widehat{E}r_{2}}{n^4} | \psi_0 \rangle \\
        & = \mathit{cst_{1}} + \frac{\mathit{cst_{2}}}{n} + \frac{\mathit{cst_{3}}}{n^2} + \frac{\mathit{cst_{2}}^*}{n} + \frac{\mathit{cst_{4}}}{n^2} + \frac{\mathit{cst_{5}}}{n^3} + \frac{\mathit{cst_{3}}^*}{n^2} \\
        & + \frac{\mathit{cst_{5}}^*}{n^3} + \frac{\mathit{cst_{6}}}{n^4} \\
        & = \mathit{cst_{1}} + \frac{2}{n}\mathrm{Re}[\mathit{cst_{2}}] + \frac{2}{n^2}\mathrm{Re}[\mathit{cst_{3}}] + \frac{\mathit{cst_{4}}}{n^2} + \frac{2}{n^3}\mathrm{Re}[\mathit{cst_{5}}] + \frac{\mathit{cst_{6}}}{n^4}
    \end{aligned}
    \end{equation*} }
    with $ \mathit{cst_{1}}, \mathit{cst_{4}}, \mathit{cst_{6}} \in \mathbb{R}, \mathit{cst_{2}}, \mathit{cst_{3}}, \mathit{cst_{5}} \in \mathbb{C} $:
    \begin{equation*}
    \begin{aligned}
        & \mathit{cst_{1}} = \langle \psi_0 | e^{-i t \widehat{H}}\widehat{H}e^{-i t \widehat{H}} | \psi_0 \rangle = \langle \psi_0 |\widehat{H}| \psi_0 \rangle\\
        & \mathit{cst_{2}} = \langle \psi_0 | e^{-i t \widehat{H}}\widehat{H}\widehat{E}r_{1} | \psi_0 \rangle\\
        & \mathit{cst_{3}} = \langle \psi_0 | e^{-i t \widehat{H}}\widehat{H}\widehat{E}r_{2} | \psi_0 \rangle\\
        & \mathit{cst_{4}} = \langle \psi_0 | \widehat{E}r_{1}^\dagger\widehat{H} \widehat{E}r_{1}| \psi_0 \rangle = |\widehat{E}r_{1}|^2 \langle Er_{1}|\widehat{H}|Er_{1} \rangle \in \mathbb{R}\\
        & \mathit{cst_{5}} = \langle \psi_0 | \widehat{E}r_{1}^\dagger\widehat{H} \widehat{E}r_{2}| \psi_0 \rangle\\
        & \mathit{cst_{6}} = \langle \psi_0 | \widehat{E}r_{2}^\dagger\widehat{H} \widehat{E}r_{2}| \psi_0 \rangle\\
    \end{aligned}
    \end{equation*}
    \Cref{fig_qc_hs_fit_energy.png} derives from \ac{neeqma} application with this observable.
    This observable is used in \cite{zhao_making_2023}.
    \item The phase is a particularly interesting observable for reconstructing the convergence of the error, as it allows for the isolation of its real and imaginary parts thanks to the Hadamard test.
    On top of giving us direct information on real and imaginary parts of Trotter error, it only needs to evaluate one single quantum circuit:
    % \subitem Order 1:
    % \begin{equation}
    % \begin{aligned}
    %     & | \langle \psi_i | e_n^{-i t \widehat{H}} | \psi_i \rangle |^2\\
    %     & \approx | \langle \psi_i | e^{-i t \widehat{H}} - \frac{2}{n} \widehat{E}r_{1} | \psi_i \rangle |^2\\
    %     & = | \langle \psi_i | e^{-i t \widehat{H}} | \psi_i \rangle - \frac{2}{n} \langle \psi_i | \widehat{E}r_{1} | \psi_i \rangle |^2\\
    %     & = | \mathit{cst_{1}} + \frac{\mathit{cst_{2}}}{n} |^2\\
    %     & \mathit{cst_{1}} \in \mathbb{R}, \mathit{cst_{2}} \in \mathbb{C}\\
    %     & \mathit{cst_{1}} = \langle \psi_i | e^{-i t \widehat{H}} | \psi_i \rangle\\
    %     & \mathit{cst_{2}} = (-2) \langle \psi_i | \widehat{E}r_{1} | \psi_i \rangle\\
    %     & Reciprocal = \frac{\mathit{cst_{2}}}{\mathit{cst_{1}}-\sqrt{n}}
    % \end{aligned}
    % \end{equation}
    % \subitem Order 2:
    \begin{equation*}
    \begin{aligned}
        & | \langle \psi_i | e_n^{-i t \widehat{H}} | \psi_i \rangle |^2\\
        & \approx | \langle \psi_i | e^{-i t \widehat{H}} + \frac{\widehat{E}r_{1}}{n} + \frac{\widehat{E}r_{2}}{n^2}| \psi_i \rangle |^2\\
        & = | \langle \psi_i | e^{-i t \widehat{H}} | \psi_i \rangle + \langle \psi_i | \frac{\widehat{E}r_{1}}{n} | \psi_i \rangle + \langle \psi_i | \frac{\widehat{E}r_{2}}{n^2}| \psi_i \rangle |^2 \\
        & = | \mathit{cst_{1}} + \frac{\mathit{cst_{2}}}{n} + \frac{\mathit{cst_{3}}}{n^2} |^2
    \end{aligned}
    \end{equation*}
    with $ \mathit{cst_{1}} \in \mathbb{R}, \mathit{cst_{2}}, \mathit{cst_{3}} \in \mathbb{C} $:
    \begin{equation*}
    \begin{aligned}
        & \mathit{cst_{1}} = \langle \psi_i | e^{-i t \widehat{H}} | \psi_i \rangle\\
        & \mathit{cst_{2}} = \langle \psi_i | \widehat{E}r_{1} | \psi_i \rangle\\
        & \mathit{cst_{3}} = \langle \psi_i | \widehat{E}r_{2} | \psi_i \rangle\\
    \end{aligned}
    \end{equation*}
    \Cref{fig_qc_hs_fit_phase.png} derives from \ac{neeqma} application with this observable.
    % \subitem Order 1 (real part):
    % \begin{equation}
    % \begin{aligned}
    %     & \mathrm{Re}[\langle \psi_i | e_n^{-i t \widehat{H}} | \psi_i \rangle ]\\
    %     & \approx \mathrm{Re}[\langle \psi_i | e^{-i t \widehat{H}} - \frac{2}{n} \widehat{E}r_{1} | \psi_i \rangle]\\
    %     & = \mathrm{Re}[\langle \psi_i | e^{-i t \widehat{H}} | \psi_i \rangle] - \frac{2}{n} \mathrm{Re}[\langle \psi_i | \widehat{E}r_{1} | \psi_i \rangle]\\
    %     & = \mathit{cst_{1}} + \frac{\mathit{cst_{2}}}{n}\\
    %     & \mathit{cst_{1}}, \mathit{cst_{2}} \in \mathbb{R}\\
    %     & \mathit{cst_{1}} = \mathrm{Re}[\langle \psi_i | e^{-i t \widehat{H}} | \psi_i \rangle]\\
    %     & \mathit{cst_{2}} = (-2) \mathrm{Re}[\langle \psi_i | \widehat{E}r_{1} | \psi_i \rangle]\\
    % \end{aligned}
    % \end{equation}
    % \subitem Order 2 (real part):
    \subitem Real part:
    \begin{equation*}
    \begin{aligned}
        & \mathrm{Re}[\langle \psi_i | e_n^{-i t \widehat{H}} | \psi_i \rangle]\\
        & \approx \mathrm{Re}[\langle \psi_i | e^{-i t \widehat{H}} + \frac{\widehat{E}r_{1}}{n} + \frac{\widehat{E}r_{2}}{n^2}| \psi_i \rangle]\\
        & = \mathrm{Re}[\langle \psi_i | e^{-i t \widehat{H}} | \psi_i \rangle] + \mathrm{Re}[\langle \psi_i | \frac{\widehat{E}r_{1}}{n} | \psi_i \rangle] \\
        & \qquad \qquad \qquad + \mathrm{Re}[\langle \psi_i | \frac{\widehat{E}r_{2}}{n^2}| \psi_i \rangle]\\
        & = \mathit{cst_{1}} + \frac{\mathit{cst_{2}}}{n} + \frac{\mathit{cst_{3}}}{n^2}
    \end{aligned}
    \end{equation*}
    with $ \mathit{cst_{1}}, \mathit{cst_{2}}, \mathit{cst_{3}} \in \mathbb{R} $:
    \begin{equation*}
    \begin{aligned}
        & \mathit{cst_{1}} = \mathrm{Re}[\langle \psi_i | e^{-i t \widehat{H}} | \psi_i \rangle]\\
        & \mathit{cst_{2}} = \mathrm{Re}[\langle \psi_i | \widehat{E}r_{1} | \psi_i \rangle]\\
        & \mathit{cst_{3}} = \mathrm{Re}[\langle \psi_i | \widehat{E}r_{2} | \psi_i \rangle]\\
    \end{aligned}
    \end{equation*}
    \Cref{fig_hs_fit_mosaique} derives from \ac{neeqma} application with this observable.
    % \subitem Order 1 (imaginary part):
    % \begin{equation}
    % \begin{aligned}
    %     & \mathrm{Im}[\langle \psi_i | e_n^{-i t \widehat{H}} | \psi_i \rangle ]\\
    %     & \approx \mathrm{Im}[\langle \psi_i | e^{-i t \widehat{H}} - \frac{2}{n} \widehat{E}r_{1} | \psi_i \rangle]\\
    %     & = \mathrm{Im}[\langle \psi_i | e^{-i t \widehat{H}} | \psi_i \rangle] - \frac{2}{n} \mathrm{Im}[\langle \psi_i | \widehat{E}r_{1} | \psi_i \rangle]\\
    %     & = \mathit{cst_{1}} + \frac{\mathit{cst_{2}}}{n}\\
    %     & \mathit{cst_{1}}, \mathit{cst_{2}} \in \mathbb{R}\\
    %     & \mathit{cst_{1}} = \mathrm{Im}[\langle \psi_i | e^{-i t \widehat{H}} | \psi_i \rangle]\\
    %     & \mathit{cst_{2}} = (-2) \mathrm{Im}[\langle \psi_i | \widehat{E}r_{1} | \psi_i \rangle]\\
    % \end{aligned}
    % \end{equation}
    % \subitem Order 2(imaginary part):
    \subitem Imaginary part:
    \begin{equation*}
    \begin{aligned}
        & \mathrm{Im}[\langle \psi_i | e_n^{-i t \widehat{H}} | \psi_i \rangle]\\
        & \approx \mathrm{Im}[\langle \psi_i | e^{-i t \widehat{H}} + \frac{\widehat{E}r_{1}}{n} + \frac{\widehat{E}r_{2}}{n^2}| \psi_i \rangle]\\
        & = \mathrm{Im}[\langle \psi_i | e^{-i t \widehat{H}} | \psi_i \rangle] + \mathrm{Im}[\langle \psi_i | \frac{\widehat{E}r_{1}}{n} | \psi_i \rangle] \\
        & \qquad \qquad \qquad + \mathrm{Im}[\langle \psi_i | \frac{\widehat{E}r_{2}}{n^2}| \psi_i \rangle] \\
        & = \mathit{cst_{1}} + \frac{\mathit{cst_{2}}}{n} + \frac{\mathit{cst_{3}}}{n^2}
    \end{aligned}
    \end{equation*}
    with $ \mathit{cst_{1}}, \mathit{cst_{2}}, \mathit{cst_{3}} \in \mathbb{R} $:
    \begin{equation*}
    \begin{aligned}
        & \mathit{cst_{1}} = \mathrm{Im}[\langle \psi_i | e^{-i t \widehat{H}} | \psi_i \rangle]\\
        & \mathit{cst_{2}} = \mathrm{Im}[\langle \psi_i | \widehat{E}r_{1} | \psi_i \rangle]\\
        & \mathit{cst_{3}} = \mathrm{Im}[\langle \psi_i | \widehat{E}r_{2} | \psi_i \rangle]
    \end{aligned}
    \end{equation*}
    \Cref{fig_hs_fit_mosaique} derives from \ac{neeqma} application with this observable.
\end{itemize}

\begin{figure}[tb]
    \begin{center}
    \resizebox{\linewidth}{!}{\includegraphics{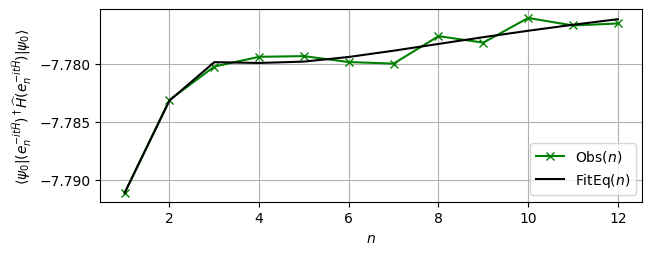}}
    \end{center}
    \caption{Observable measurement (energy of the Trotterized Hamiltonian simulation) at different Trotter numbers $n$ with the associated error model, with the free-parameters adjusted by classical optimization.
    This curve was computed at Hamiltonian simulation's time $ t = 1 $.}
    \label{fig_qc_hs_fit_energy.png}
\end{figure}

\begin{figure}[tb]
    \begin{center}
    \resizebox{\linewidth}{!}{\includegraphics{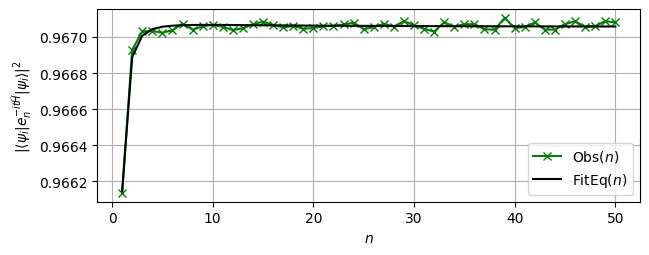}}
    \end{center}
    \caption{Observable measurement (phase of the Trotterized Hamiltonian simulation) at different Trotter numbers $n$ with the associated error model, with the free-parameters adjusted by classical optimization.
    This curve was computed at Hamiltonian simulation's time $ t = 1 $.}
    \label{fig_qc_hs_fit_phase.png}
\end{figure}

\subsection{Related Work}
This section gives an overview of the related work \cite{ikeda_measuring_2024} % and \cite{mehendale_estimating_2025}
previously mentioned.
In order to underline the differences with our approach, \cite{ikeda_measuring_2024}'s algorithm: Trotter24 is presented using a workflow with a structure similar to the one used for \ac{neeqma} \Cref{workflow_neeqma}.

Trotter24 aims at realizing the Hamiltonian simulation under the constraint of a defined accuracy at each time step. The way the autors proceed to reduce their Trotter error is by starting with the simulation of a chosen Hamiltonian on a small time step and estimate the Trotter error (the estimated observable is the fidelity, as described in \Cref{section_fidelity}). If the error is greater than $\epsilon$ (desired accuracy), Trotter24 restarts the simulation with an even smaller time step until the error is less than $\epsilon$. From there, it repeats the process until it has completely simulated the Hamiltonian (when the sum of the saved time steps reaches the time t). Since the sizes of the selected time steps can be different, this iterative adaptive algorithm proposal cannot be considered equivalent to working with a Trotter number.

\begin{figure}[tb]
\begin{center}
\resizebox{\linewidth}{!}{\includegraphics{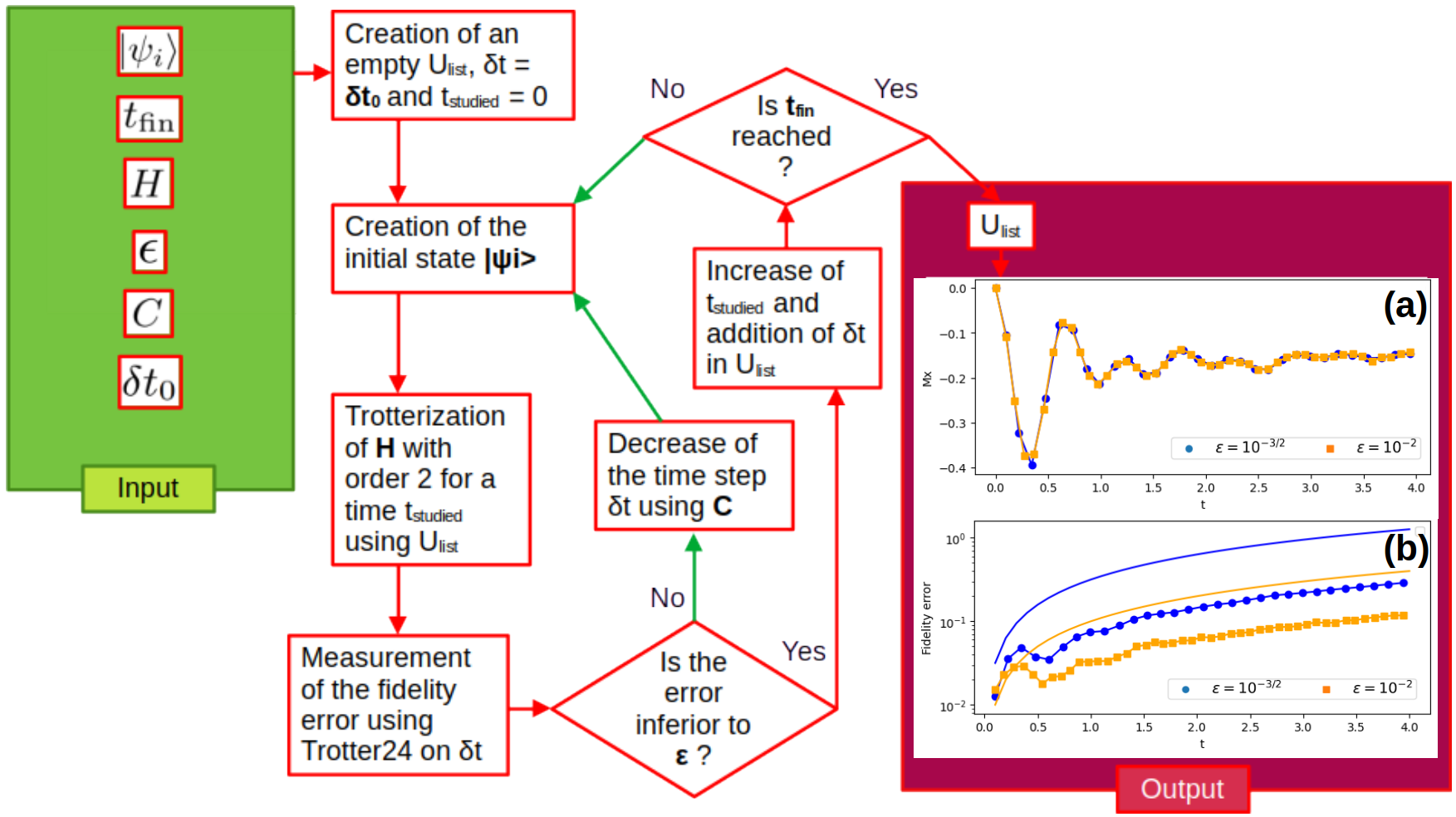}}
\end{center}
\caption{Trotter24 workflow \cite{ikeda_measuring_2024}.
\textbf{(a)} Dynamics of x-magnetization density calculated by fidelity-based Trotter24 for tolerance $\epsilon = 10^{-3/2}$ (circle) and $10^{-2}$ (square).
\textbf{(a)} is a reproduction of \cite[fig 2.a]{ikeda_measuring_2024}.
\textbf{(b)} Fidelity errors in the simulation presented in \textbf{(a)}.
Solid curves show upper bounds. Blue (orange) points and curve correspond to $\epsilon = 10^{-3/2}$ ($10^{-2}$).
\textbf{(b)} is a reproduction of \cite[fig 2.b]{ikeda_measuring_2024}.
}
\label{fig:ikeda-workflow}
\end{figure}

This section presents the details of the algorithm realized in the related work \cite{ikeda_measuring_2024} using a workflow \Cref{fig:ikeda-workflow}. The
Workflow description follows:
\begin{itemize}

\item As inputs it is required to have an initial state $|\psi_i \rangle$, a total simulation time $t_{\mathit{fin}}$, a problem $\widehat{H}$, a time step error bound $\epsilon$, a constant $C \in [0, 1]$ used to update the time step duration, an $\emph{initial time step duration}$ $\delta t_0$.

\item It starts by creating $U_{\mathit{list}}$ to save time steps that respect the accuracy constraint $\epsilon$, it also fixes the $\emph{actual time step duration}$ $\delta_t$ to $\delta t_0$ and the $\emph{studied time}$ to 0.

\item After initializing $|\psi_i \rangle$ in a quantum circuit, it applies a series of Trotter gates on $\widehat{H}$ with order 2 for every time steps saved in $U_{\mathit{list}}$ (when $U_{\mathit{list}}$ is empty, it simply skips this step).

\item Using the time step $\delta_t$, it applies a Trotter gate of order 2, followed by an adjoint Trotter gate of order 4 on the same time step $\delta_t$ and on the same Hamiltonian $\widehat{H}$.
% It is similar to a fidelity measurement, if there is no gate error, the two gates are equals and the fidelity read $1$ \Cref{section_fidelity}. % This corresponds to the blue and red sections of \Cref{fig:ikeda-measurement-method}

\item To cancel out unnecessary terms (time steps saved in $U_{\mathit{list}}$), it applies a series of adjoints Trotter gates on $\widehat{H}$ with order 2 for every time steps saved in $U_{\mathit{list}}$ (when $U_{\mathit{list}}$ is empty it simply skips this step).
It then measures all qubits and calculates the fidelity error.

\item If the fidelity error is greater than $\epsilon$, it restarts the process with a lower time step: $ \delta_t \longleftarrow C \delta_t $.

\item If the fidelity error is lower than $\epsilon$, it saves $\delta_t$ in $U_{\mathit{list}}$, increases $t_{\mathit{studied}}$ with $\delta_t$.

\item If the total time $t_{\mathit{fin}}$ is not yet reached, it starts again the quantum circuit, this time using the updated $U_{\mathit{list}}$.

\item If the total time $t_{\mathit{fin}}$ is reached, it outputs $U_{\mathit{list}}$ containing all time step durations required to perform Trotterization on a given time $t_{\mathit{fin}}$ while respecting the constraint given by $\epsilon$ on any time steps.

% \begin{figure}[tb]
% \begin{center}
% \resizebox{\linewidth}{!}{\includegraphics{fig_expe/ikeda-magnetization.png}}
% \end{center}
% \caption{Dynamics of x-magnetization density calculated by fidelity-based Trotter24 for tolerance $\epsilon = 10^{-3/2}$ (circle) and $10^{-2}$ (square).}
% \label{fig:ikeda-magnetization}
% \end{figure}

% \begin{figure}[tb]
% \begin{center}
% \resizebox{\linewidth}{!}{\includegraphics{fig_expe/ikeda-fidelity-error.png}}
% \end{center}
% \caption{Fidelity errors in the simulation presented in \Cref{fig:ikeda-magnetization}. Solid curves show upper bounds. Blue (orange) points and curve correspond to $\epsilon = 10^{-3/2}$ ($10^{-2}$).}
% \label{fig:ikeda-fidelity-error}
% \end{figure}

\item To check whether the simulation is accurate, the authors measure in the Z-basis at each time step, the x-magnetization $\widehat{m_x} \equiv \frac{1}{L} \sum_{j=1}^{L} \widehat{\sigma_j^x}$ of the Trotterization shown on \Cref{fig:ikeda-workflow} \textbf{(a)}, since the x-magnetization evolution follows a natural and expected sinusoidal curve, it shows an accurate simulation.

\item Finally, the authors have plotted the accumulated error over the whole simulation (point curves) represented by \Cref{fig:ikeda-workflow} \textbf{(b)}.
To calculate the accumulated error, Trotter24 initializes $|\psi_i \rangle$ in a quantum circuit, and apply a series of Trotter gates on $\widehat{H}$ with order 2 for every time steps saved in $U_{\mathit{list}}$ followed by a series of adjoints Trotter gates on $\widehat{H}$ with order 4 for every time steps saved in the reversed $U_{\mathit{list}}$.
The upper bound curves are calculated assuming fidelity errors equal $\epsilon$ on every time steps in $U_{\mathit{list}}$, this can be achieved when $C \approx 1$.
\end{itemize}

\subsection{Experimental Details}
All experiments were realized using quantum circuits with a number of shots equal to $10^{5}$, except the experiments to measure $Obs(n)$ in \Cref{fig_qc_hs_fit_phase.png} and \Cref{fig_hs_fit_mosaique} where we used $10^{8}$ shots.
\subsubsection{Optimizer Parameters}
For the Hamiltonian Simulation experiment, we have used the \SaveVerb{verbatimtext} 'curve_fit' \UseVerb{verbatimtext} method from the \SaveVerb{verbatimtext} 'scipy.optimize' \UseVerb{verbatimtext} package.
For the QSP experiment, we have used the \SaveVerb{verbatimtext} 'cobyla' \UseVerb{verbatimtext} method from the \SaveVerb{verbatimtext} 'scipy.optimize' \UseVerb{verbatimtext} package.
\subsubsection{Sign Function Phase Angles}
Each list contains a size associated with a different degree.
These lists are obtained thanks to the prompt 
\begin{verbatim}
'pyqsp --plot --polyargs=1,d --plot-real-only 
--polyname poly_sign poly --return-angles'
\end{verbatim}
in the \cite[pyqsp code]{chao_finding_2020, dong_efficient_2021, martyn_grand_2021} package, all the raw phase angles are contained in the associated folder
\begin{verbatim}
'sign_function_qsp_phase_angle_from_pyqsp.txt'.
\end{verbatim}
\subsubsection{Hamiltonian}
Utilized normalization factor (obtained from numpy): $ |\lambda_{m}| = 8.654853588861483 $.
$LiH$ Hamiltonian Pauli-string decomposition is contained in the associated file named \SaveVerb{verbatimtext} 'lih_2a.txt' \UseVerb{verbatimtext}.

\newpage
\begin{figure*}[htb]
\centering
\begin{equation}
\begin{aligned}
  \text{Starting with: } &
    \left\{
    \begin{array}{ll}
        \widehat{C}(\widehat{A}, \widehat{B}) & = \widehat{A} + \widehat{B} + \frac{[\widehat{A}, \widehat{B}]}{2} + \frac{[\widehat{A} [\widehat{A}, \widehat{B}]] + [\widehat{B}, [\widehat{B}, \widehat{A}]]}{12} \\
        \widehat{B} & = t \widehat{H_{N + 1}} \\
        \widehat{A}_{N} & = t \sum_{j = 0}^{N} \widehat{H_{j}} + t^{2} \widehat{\Xi_{1, N}} + t^{3} \widehat{\Xi_{2, N}} + \mathcal{O}(t^{4}) % \\
        % & = t \sum_{j = 0}^{N} \widehat{H_{j}} + t^{2} \widehat{\Xi_{1, N}} + \mathcal{O}(t^{3}) \\
        % & = t \sum_{j = 0}^{N} \widehat{H_{j}} + \mathcal{O}(t^{2})
    \end{array}
    \right. \\
    \text{we want to show that: } &
    \left\{
    \begin{array}{ll}
        \widehat{\Xi_{1, N}} & = \frac{1}{2} \sum_{0 \leqslant j < k \leqslant N} [\widehat{H_{j}}, \widehat{H_{k}}] \\
        \widehat{\Xi_{2, N}} & = \frac{1}{12} ( \sum_{j = 0}^{N} \sum_{k = 0}^{N} [\widehat{H_{j}}, [\widehat{H_{j}}, \widehat{H_{k}}]] + 2 \sum_{0 \leqslant j < k < l \leqslant N} ( [\widehat{H_{j}}, [\widehat{H_{k}}, \widehat{H_{l}}]] + [\widehat{H_{l}}, [\widehat{H_{k}}, \widehat{H_{j}}]] ))
    \end{array}
    \right. \\
    \text{For $ N = 0 $: } & \widehat{A}_{0} =  t \sum_{j = 0}^{0} \widehat{H_{j}} + t^{2} \widehat{\Xi_{1, N=0}} + t^{3} \widehat{\Xi_{2, N=0}} = t \widehat{H_{0}} \text{ which is true.} \\
    \text{Recurrence: } & \widehat{A}_{N + 1} = \widehat{C}(\widehat{A}_{N}, \widehat{B}) \\
    & \quad \qquad = t \sum_{j = 0}^{N} \widehat{H_{j}} + t^{2} \widehat{\Xi_{1, N}} + t^{3} \widehat{\Xi_{2, N}} + t \widehat{H_{N + 1}} + \frac{[(t \sum_{j = 0}^{N} \widehat{H_{j}} + t^{2} \widehat{\Xi_{1, N}}), t \widehat{H_{N + 1}}]}{2} \\
    & \qquad \qquad \qquad + \frac{[t \sum_{j = 0}^{N} \widehat{H_{j}} [t \sum_{j = 0}^{N} \widehat{H_{j}}, t \widehat{H_{N + 1}}]] + [t \widehat{H_{N + 1}}, [t \widehat{H_{N + 1}}, t \sum_{j = 0}^{N} \widehat{H_{j}}]]}{12} + \mathcal{O}(t^{4}) \\
    & \quad \qquad = t \sum_{j = 0}^{N + 1} \widehat{H_{j}} + t^{2} \widehat{M_{1}} + t^{3} \widehat{M_{2}} + \mathcal{O}(t^{4}) \\
    \text{with: } & \widehat{M_{1}} = \widehat{\Xi_{1, N}} + \frac{[\sum_{j = 0}^{N} \widehat{H_{j}}, \widehat{H_{N + 1}}]}{2} = \sum_{0 \leqslant j < k \leqslant N + 1} \frac{[\widehat{H_{j}}, \widehat{H_{k}}]}{2} = \widehat{\Xi_{1, N+1}} \\
    \text{and: } & \widehat{M_{2}} = \widehat{\Xi_{2, N}} + \sum_{0 \leqslant j < k \leqslant N} \frac{[[\widehat{H_{j}}, \widehat{H_{k}}], \widehat{H_{N + 1}}]}{4} + \frac{\sum_{j = 0}^{N} \sum_{k = 0}^{N} [\widehat{H_{j}} [\widehat{H_{k}}, \widehat{H_{N + 1}}]] + \sum_{j = 0}^{N} [\widehat{H_{N + 1}}, [\widehat{H_{N + 1}}, \widehat{H_{j}}]]}{12} \\
    & \quad \qquad = \frac{1}{12} ( \sum_{j = 0}^{N} \sum_{k = 0}^{N} [\widehat{H_{j}}, [\widehat{H_{j}}, \widehat{H_{k}}]] + 2 \sum_{0 \leqslant j < k < l \leqslant N} ( [\widehat{H_{j}}, [\widehat{H_{k}}, \widehat{H_{l}}]] + [\widehat{H_{l}}, [\widehat{H_{k}}, \widehat{H_{j}}]] ) \\
    & \qquad \qquad \qquad + 3 \sum_{0 \leqslant j < k \leqslant N} [[\widehat{H_{j}}, \widehat{H_{k}}], \widehat{H_{N + 1}}] + \sum_{j = 0}^{N} \sum_{k = 0}^{N} [\widehat{H_{j}} [\widehat{H_{k}}, \widehat{H_{N + 1}}]] + \sum_{j = 0}^{N} [\widehat{H_{N + 1}}, [\widehat{H_{N + 1}}, \widehat{H_{j}}]] ) \\
    \text{Using: } \\
    (*) & \sum_{j = 0}^{N} \sum_{k = 0}^{N} [\widehat{H_{j}} [\widehat{H_{k}}, \widehat{H_{N + 1}}]] = \sum_{j = 0}^{N} [\widehat{H_{j}} [\widehat{H_{j}}, \widehat{H_{N + 1}}]] + \sum_{0 \leqslant j < k \leqslant N} ( [\widehat{H_{j}} [\widehat{H_{k}}, \widehat{H_{N + 1}}]] + [\widehat{H_{k}} [\widehat{H_{j}}, \widehat{H_{N + 1}}]] ) \\
    & \Rightarrow \widehat{M_{2}} = \frac{1}{12} ( \sum_{j = 0}^{N+1} \sum_{k = 0}^{N+1} [\widehat{H_{j}}, [\widehat{H_{j}}, \widehat{H_{k}}]] + 2 \sum_{0 \leqslant j < k < l \leqslant N} ( [\widehat{H_{j}}, [\widehat{H_{k}}, \widehat{H_{l}}]] + [\widehat{H_{l}}, [\widehat{H_{k}}, \widehat{H_{j}}]] ) \\
    & \qquad \qquad \qquad + 3 \sum_{0 \leqslant j < k \leqslant N} [\widehat{H_{N + 1}}, [\widehat{H_{k}}, \widehat{H_{j}}]] + \sum_{0 \leqslant j < k \leqslant N} ( [\widehat{H_{j}} [\widehat{H_{k}}, \widehat{H_{N + 1}}]] + [\widehat{H_{k}} [\widehat{H_{j}}, \widehat{H_{N + 1}}]] )) \\
    \text{Using: } \\
    (**) & 0 = [k, [j, l]] + [j, [l, k]] + [l, [k, j]] \\
    & \Rightarrow \widehat{M_{2}} = \frac{1}{12} ( \sum_{j = 0}^{N+1} \sum_{k = 0}^{N+1} [\widehat{H_{j}}, [\widehat{H_{j}}, \widehat{H_{k}}]] + 2 \sum_{0 \leqslant j < k < l \leqslant N} ( [\widehat{H_{j}}, [\widehat{H_{k}}, \widehat{H_{l}}]] + [\widehat{H_{l}}, [\widehat{H_{k}}, \widehat{H_{j}}]] ) \\
    & \qquad \qquad \qquad + 2 \sum_{0 \leqslant j < k \leqslant N} ( [\widehat{H_{N + 1}}, [\widehat{H_{k}}, \widehat{H_{j}}]] + [\widehat{H_{j}} [\widehat{H_{k}}, \widehat{H_{N + 1}}]])) \\
    & \quad \qquad = \widehat{\Xi_{2, N+1}} \\
\end{aligned}
\label{eq_xi_proof}
\end{equation}
\caption{$\widehat{\Xi}$ formula proof by induction.}
\end{figure*}

\end{document}